\documentclass{aa}
\usepackage{upgreek}
\usepackage[varg]{txfonts}
\usepackage{array}
\newcolumntype{R}{>{\raggedright\arraybackslash}p{9.2cm}}
\newcolumntype{S}{>{\raggedright\arraybackslash}p{6.5cm}}
\usepackage{enumitem}

\begin{document}

\title{Simulations of a 2~$\times$~1.5D coded aperture camera \\
  for X-ray astronomy}

\author{
J.J.M. in `t Zand\inst{1}
\and
L. Kuiper\inst{1}
\and
F. Ceraudo\inst{2}
\and
Y. Evangelista\inst{2}
\and
M. Hernanz\inst{3,4}
\and
A. Patruno\inst{3}
}

\institute{SRON Space Research Organization Netherlands, Niels
Bohrweg 4, 2333 CA Leiden, the Netherlands
\and
INAF-IAPS Roma, via Fosso del Cavaliere 100, 00133 Rome, Italy
\and
Institute of Space Sciences (ICE-CSIC), Campus UAB, 08193 Cerdanyola
del Vallès (Barcelona), Spain
\and
Institut d'Estudis Espacials de Catalunya (IEEC), Barcelona, Spain
}

\abstract{ The concept of two perpendicular one-dimensional coded
  aperture cameras, necessitated by the imaging capability of the
  detector, finds its application in the design of the Wide Field
  Monitor (WFM). This instrument has the future goal to monitor the
  variable X-ray sky for transient activity. Characteristic of each
  camera is a fine angular resolution in one direction (typically 5
  arcmin) and a coarse one in the other (5 degrees). The coarse
  perpendicular resolution makes the camera so-called '1.5D'. The WFM
  has been studied for a number of space-borne X-ray observatory
  concepts: LOFT, eXTP, Strobe-X, ARCO and now LEM-X. We here report
  on a study of two decoding algorithms for this instrument and its
  imaging performance. Detector responses to the X-ray sky are
  simulated, including the signal processing by the front-end and
  back-end electronics. The decoding algorithms are the iterative
  removal of sources (IROS), in combination with cross correlation,
  and the maximum likelihood method (MLM). IROS is most suited for the
  determination of the point source configuration of the observed sky
  and MLM for the optimum determination of the source fluxes. The
  simulation results show that despite the 1.5D imaging of each
  camera, the reconstruction of scientific data is as if each camera
  pair were replaced by a single 2D camera with the same specification
  in angular resolution and a detector size equal to that of the two
  in the pair, except that source confusion is somewhat larger in the
  former.  Thus, the WFM is a high performance monitoring instrument
  with straightforward and proven technology that enables the
  identification of new cosmic X-ray sources, for instance X-ray
  novae, gamma-ray bursts and electromagnetic counterparts to
  gravitational wave events from merging compact objects, and the
  detection of unusual and interesting behavior of persistent cosmic
  X-ray sources, such as accretion disk state changes.}

\keywords{Astronomical instrumentation: miscellaneous - Methods: numerical -
Techniques: image processing - telescopes}

\maketitle
\nolinenumbers

\section{Introduction}
\label{intro}

Apart from rotation, the sky seems static by the naked eye but is by
far not. Particularly in X-rays, it is vibrant with variability and
transient activity on all time scales. This became clear already early
on in the 1960s at the dawn of X-ray astronomy, through
instrumentation specifically designed for detecting transient activity
on Earth (atmospheric nuclear weapons tests) through the Vela
satellites \citep[e.g.,][]{klebesadel1973}. This motivated the
implementation of a long series of X-ray all-sky monitors (ASMs)
throughout history, the most recent examples being the Monitoring
All-sky X-ray Imager MAXI launched in 2009 to the International Space
Station \citep[ISS;][]{matsuoka2009}, the cubesats Ninjasat
\citep{tamagawa2025} and Blackcat \citep{falcone2024}, the Wide Field
X-ray Telescope (WXT) on the Einstein Probe (EP) launched in 2024
\citep{yuan2022} and ECLAIRs on the Space Variable Objects
Monitor \citep[SVOM;][]{godet2014,givaudan2024} also launched in
2024. This will continue in the future. The modernization of the ASM
fleet concentrates on the extension of the bandpass to longer
wavelengths than the classical X-ray regime, i.e. $>$1 nm, and higher
sensitivity and higher duty cycles, i.e. larger field of views, thus
collecting larger exposure times per point on the sky per
day. Reaching larger fields of view is accomplished by either coded
aperture imaging, a technique applied since the late 1970s and
currently in wide use through for instance BAT on Swift
\citep{barthelmy2005} and ECLAIRs, time multiplexing, such as in
MAXI, and Lobster eye optics, a technique applied since only
since a few years through LEIA \citep{zhang2022} and WXT. Coded
aperture imaging \citep{dicke1968,ables1968} is a relatively simple
and low-cost technique. Although it is less sensitive than direct
focusing techniques by a few orders of magnitude, it is sensitive
enough to detect many interesting transients in our own galaxy (e.g.,
accreting stellar black holes, thermonuclear flashes on neutron stars,
stellar flares) and beyond (e.g., GRBs). This technique has, therefore,
been very successful.

Coded aperture cameras \citep[for reviews
  see,][]{caroli1987,zand1992,skinner2008,braga2020,goldwurm2022} are
usually based on a planar detector with identical spatial resolutions
in both dimensions, yielding identical angular resolutions in those
dimensions.  This is a convenient setup. However, this is not strictly
necessary. For example, the SuperAGILE instrument on the AGILE
satellite \citep{feroci2010,evangelista2010} consisted of two pairs of
identical cameras with each camera a high resolution in only one
dimension and a coarse resolution in the other.  By co-aligning both
cameras to the same field of view but orthogonally with respect to
each other, the 1D cameras accomplish a high angular resolution in
both dimensions. This more involved setup was necessitated by the 1D
nature of the silicon microstrip detectors but also has a number of
advantages in terms of weight and power consumption. The present paper
is about an instrument concept similar to that of SuperAGILE but
involving better performing silicon drift detectors (SDDs) and more
pairs of orthogonal cameras to increase the field of view. It is
called the Wide-Field Monitor (WFM). While other WFM publications
focus on the hardware implementation of this instrument, we here focus
on the principles of the software implementation: the development of
an image and spectral reconstruction algorithm customized for the pair
of coded aperture camera setup, the investigation of the performance
with these algorithms and a comparison with a would-be 2D setup. This
paper focuses on principles and first-order assessments of performance
and as such is a state-of-the-art report. In the future, software
development needs to continue and simulations need to reach a higher
level of fidelity, ultimately culminating in analysis software for
real data.

After introducing the WFM in more detail in Sect.~\ref{wfm}, we
summarize decoding algorithms in general and in particular for the WFM
in Sect.~\ref{alg}. In Sect.~\ref{approach}, we explain how we
approach the simulations of the sources, background and the instrument
response. In Sect.~\ref{sims}, we discuss the simulation results of the
images and what this says about the performance of the WFM, in
Sect.~\ref{specsims} for the spectral simulations. We end in
Sect.~\ref{conc} with the conclusions of our study and the way forward
with regards to the treatment of WFM data once the instrument flies.

\section{The Wide Field Monitor}
\label{wfm}

\begin{figure}
\includegraphics[width=\columnwidth,angle=0,trim=0 0.cm 8cm 18cm,clip=true]{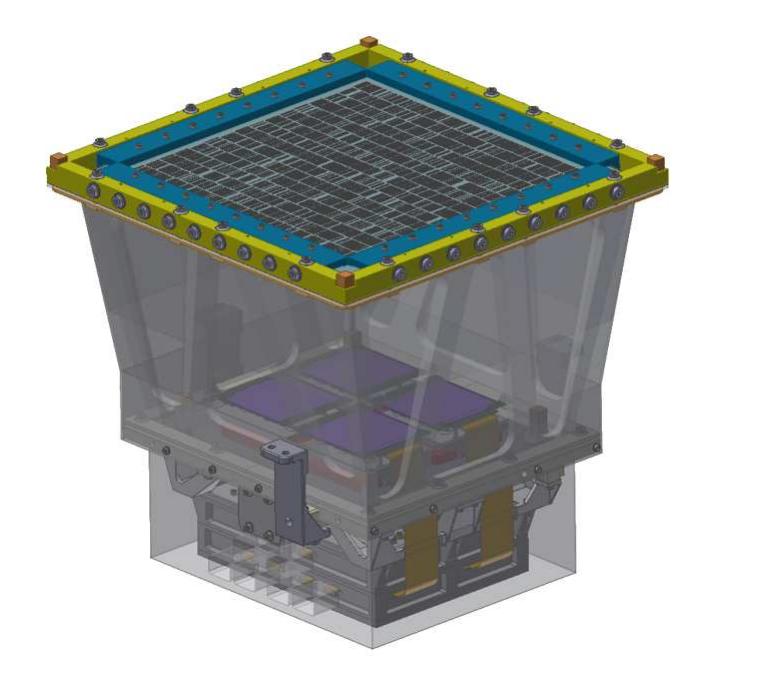}
\caption{WFM camera shown with the shielding being made slightly
  transparent. In purple, the 4 SDDs are depicted, on top the mask.}
\label{f1}
\end{figure}

The concept of the Wide Field Monitor instrument was conceived 15
years ago and found a first application in the proposal for the Large
Observatory For Timing \citep[LOFT;][]{feroci2012,brandt2014},
followed by concept studies for the enhanced X-ray Timing and
Polarimetry mission
\citep[eXTP;][]{zhang2019,hernanz2018,hernanz2022,hernanz2024,zwart2022},
the Spectroscopic Time-Resolving Observatory for Broadband X-rays
\citep[STROBE-X;][]{ray2024,remillard2024}, the Astrophysics of
Relativistic Compact Object mission (ARCO) submitted to the call for
F3 mission concepts for ESA in 2025, and the Lunar Electromagnetic
Monitor in X-rays \citep[LEM-X;][]{delmonte2024}. It is an array of
wide-field cameras for monitoring transient activity in the X-ray sky
at photon energies between 2 and 50 keV. The field of view of each
camera is 20\% of the full sky and an array, therefore, enables
exceptionally high duty cycle monitoring. The instrument consists
basically of N pairs of orthogonal 1D coded aperture cameras. They are
often referred to as 1.5D cameras because the SDDs have resolution,
albeit coarse, in the perpendicular dimension
\citep[e.g.,][]{hernanz2018}. Table~\ref{t1} summarizes the
characteristics of one WFM camera as designed for eXTP. A prototype of
such a camera is currently under construction at ICE-CSIC (Barcelona).

\begin{figure}
\includegraphics[width=\columnwidth,angle=0,trim=1cm 0cm 2cm 12cm,clip=true]{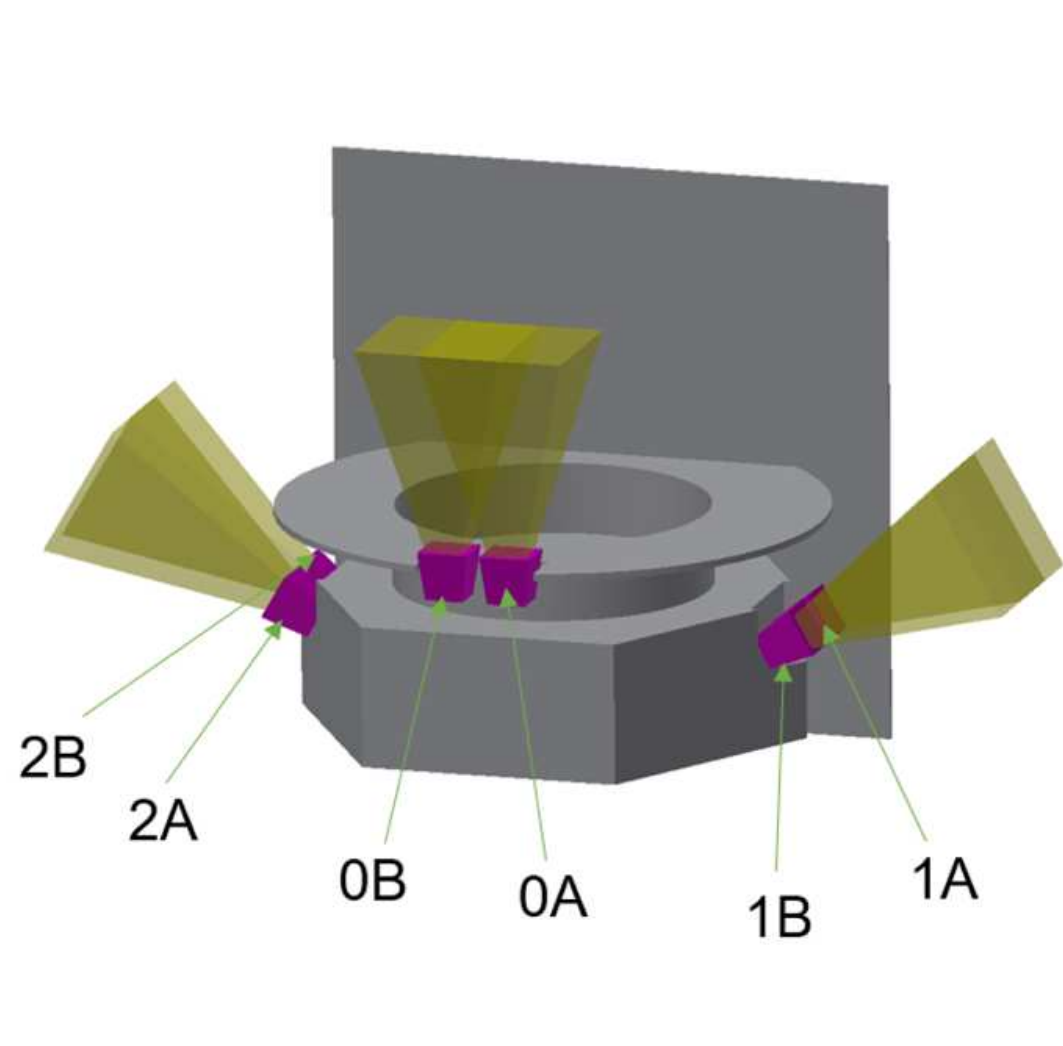}
\caption{Par example, a schematic of the WFM configuration foreseen
  for eXTP. The narrow-field instruments of eXTP are pointed upward, so
  WFM pair '0' covers the pointing of those instruments at the center of
  the field of view.}
\label{f2}
\end{figure}

\begin{table*}
\caption{Specifications of WFM relevant for imaging}
\label{t1}
\begin{tabular}{lSR}
\hline\hline
Item & Parameter & Value\\
\hline
Mask & Physical size & 260.0 mm $\times$ 260.0 mm $\times$ 0.15 mm \\
     & Pattern & 1040$\,\times\,$16 elements of 0.25$\,\times\,$16.4 mm$^2$ (the latter encompassing
       2.4 mm ribs and 14.0 mm open areas) \\
     & Material & Tungsten 0.15 mm thick ($<$0.3\% transparent for
       E$<$30 keV) \\
Detector & Type & 4 Silicon drift detectors in a 2$\,\times\,$2 array each \\
     & Size & 77\,mm\,$\times$\,72\,mm\,$\times$\, 0.45\,mm each \\
     & Boundary of combined sensitive area & 158$\,\times\,$153 mm$^2$ \\
     & Quantum efficiency & $>$97\% for E$<$10 keV,
       $>$33\% for E$<$20 keV \\
     & Position resolution (FWHM) & $<$60 µm along X and $<$8 mm along Y \\
     & Energy resolution (FWHM) & $<300$~eV at 6 keV \\
Foils & Thickness multi-layer insulator above mask & 12.70~$\upmu$m kapton + 0.10 $\upmu$m Al + 0.15 $\upmu$m SiO$_2$ \\
      & Thickness micrometeorite protection & 15~$\upmu$m polypropylene \\
Geometry & Mask-detector distance & 203.05 mm (from top mask surface to
top SDD surface) \\
     & Field of view & 90$\,\times\,$90 square degrees (29$\,\times\,$29 square degrees FCFOV) \\
     & Angular resolution (FWHM, on axis)& fine 4.4 arcmin, coarse 4.6 deg \\
Data & Photon positions & on detector full resolution (unbinned) \\
     & Time binning & 10\;$\upmu$s nominal \\
     & Photon energy binning & 1024 channels \\ 
\hline\hline
\end{tabular}
\end{table*}

Each WFM camera hosts 2$\,\times\,$2 SDDs in one detector plane that
have a high spatial resolution in one direction and a coarse
resolution in the other (see Fig.~\ref{f1}). Each of the 4 detectors
has a sensitive area of 6.5$\,\times\,$7.0 cm$^2$. The 2$\,\times\,$2
detector array is interspaced by 1.3 cm along the coarse direction and
2.8 cm along the fine direction. This space is a dead area in the
detector plane. The detector plane has sensitivity within a
15.8$\,\times\,$15.3 cm$^2$ boundary. A coded mask of size
26.0$\,\times\,$26.0 cm$^2$ is 20.29 cm above the detector plane. The
mask pattern consists of 1040$\,\times\,$16 elements of size
0.025$\,\times\,$1.64 cm$^2$, thus, after ignoring the first and last
rib (see below) covering 26.0$\,\times\,$26.0 cm$^2$ (here X is
assumed along the fine direction and Y along the coarse). The space
between mask and detector plane is shielded by a collimator to prevent
photons from reaching the detector without passing through the
mask. Each camera has a field of view of 90$\,\times\,$90 square
degrees of which the central 29$\,\times\,$29 square degrees
illuminate the full detector array (with part of the mask) and is
(somewhat confusingly) called the `fully coded' field of view.  The
optical design of the camera is basically unchanged since 2010
\citep[e.g.,][]{donnarumma2012,brandt2014}. To produce a setup with
equal high angular resolution in both dimensions, each camera is
paired with an identical camera rotated by 90 degrees along the
optical axis. The ultimate WFM configuration consists of N pairs of
cameras, each pointed at a different position on the sky. In LOFT,
N=4, in eXTP N=3 (see Fig.~\ref{f2}), in STROBE-X N=4, in ARCO N=1 and
in LEM-X N=7.

\begin{figure}[t] 
\includegraphics[width=\columnwidth,angle=0,trim=0.5cm 0cm 4.5cm 13cm,clip=true]{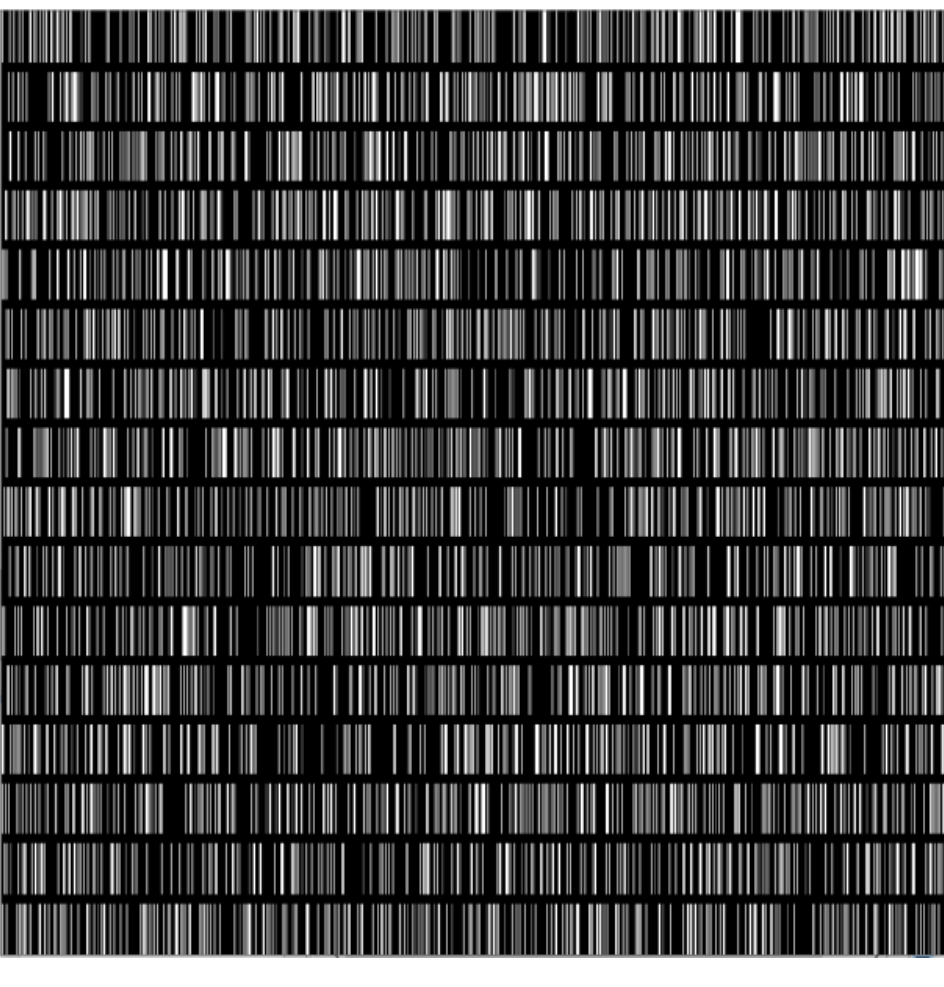}
\caption{WFM mask. The cyclic difference set is laid down starting
  from the bottom left, going first from left to right and then bottom
  to top.}
\label{f3}
\end{figure}

The mask pattern is based on a cyclic difference set from biquadratic
residues of primes $v=4x^2+1$ with $x$ odd \citep[theorem 5.16
  in][]{baumert1971}, see also \cite{gottesman1989}, where set
elements are $j=$mod($i^4$,$v$) for $i=1..v$. This yields that 25\% of
the numbers between 1 and $v$ are a member. At those positions the
mask pattern is defined to be open, at all the others closed. For the
mask in the WFM, $v$ was chosen to be $16901$ ($x=65$). This mask
pattern is folded in 2 dimensions over a 1040$\,\times\,$16 array of
0.25 mm$\,\times\,$16.4 mm elements with the last 261 members of the
biquadratic residue set not used. The mask incorporates 2.4 mm closed
ribs within the 16.4 mm long slits, so that the effective open area of
the mask becomes 21.6\%, see Fig.~\ref{f3}.

The WFM is, like all coded aperture cameras and perhaps even more so
because of its 2$\,\times\,$1.5D nature, an indirect imaging
instrument. In contrast to focusing telescopes, the signal of all
cosmic sources is coded and mixed which gives rise to the disadvantage
that there is cross talk between sources and decreased sensitivity for
the same photon-collecting area. An essential ingredient of a coded
aperture camera is the algorithm to decode the signal from the coded
detector domain to the decoded sky domain. The purpose of the present
paper is to discuss methods to treat the data through such algorithms
and verify through detailed simulations the expected performance.

\section{Algorithms to decode detector data to sky data}
\label{alg}

\begin{figure*}
\center
\includegraphics[height=1.5\columnwidth,angle=270,trim=0cm 0cm 0cm 0cm,clip=true]{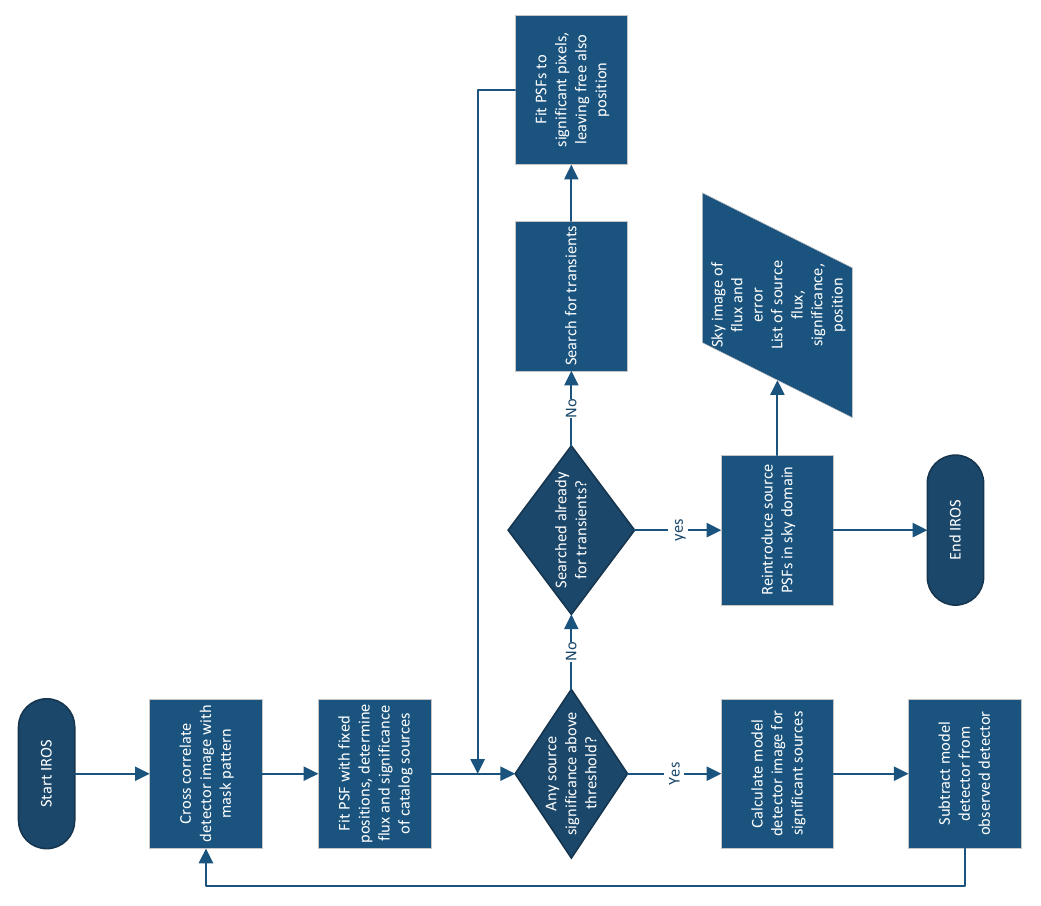}
\caption{Flow diagram of IROS.}
\label{f4}
\end{figure*}

The literature discusses various successful algorithms to accomplish
reconstruction of sky data from coded aperture detector data
\citep[e.g.,][]{rideout1995}. This includes cross correlation
\citep{brown1974,fenimore1978}, the maximum likelihood method
\citep[MLM;][]{skinner1993,nottingham1993}, the maximum entropy method
\citep{willingale1981}, Wiener filtering \citep{sims1980,rideout1996}
and machine-learning techniques \citep{meissner2023}. The cross
correlation method has been extended with iterative removal of sources
\citep[IROS][]{hammersley1986,hammersley1992,zand1992}, fine sampling
and delta decoding \citep{fenimore1981}. We here develop the methods
of IROS and MLM for application in WFM. These methods have different
purposes.  IROS excels in fast sky image reconstruction and search for
previously unknown point sources. MLM excels in the best determination
of fluxes and positions. IROS provides a map of the complete field of
view by the fast cross correlation technique using FFTs. MLM can
provide a sky map by calculating detector images for every possible
sky location which is a slower process. Our approach to data handling
of WFM data is to first find all point sources through IROS and then
characterize all those point sources (i.e., determine positions, light
curves and spectra) through MLM.

\subsection{Iterative Removal of Sources}

\subsubsection{Outline}

IROS is an analogue of the `clean' algorithm in interferometric
radio-astronomy imaging \citep{hogbom1974}. The basis of IROS
\citep{hammersley1986} is a cross correlation of the detector image
with a mathematical version of the mask pattern that ensures that the
expected flux at any pixel where there is no source is zero and is
properly normalized to phot s$^{-1}$ cm$^{-2}$ for any source
pixel. The geometrical response of the instrument needs to be taken
into account to properly normalize by cm$^2$, for instance obscuration
by support structures of the detector or mask. Any photon
energy-dependent response parameters of the instrument, such as
detector spatial resolution, mask transparency and shadowing need to
be taken into account through for instance modeling of the point
spread function in the sky and detector domains.

A typical feature of cross correlation images is `coding noise'. Mask
patterns are chosen such that their autocorrelation is a delta
function \citep{palmieri1974,fenimore1978}. This requires the
application of patterns based on pseudo-random arrays, such as cyclic
difference sets, and the proper lay-down in two
dimensions. Figure~\ref{f3} shows the mask pattern of the WFM
envisaged for eXTP. Even with a pseudo-random pattern, the
autocorrelation is only 2-valued if the full pattern is involved in
coding any sky position. There are some camera configurations where
this is accomplished, for instance by applying a repeated ideal mask
pattern and putting a collimator on the detector that limits the field
of view from any position on the detector to one complete cycle of
such a pattern \citep{hammersley1986}. However, in many practical
applications (e.g., when sensitivity or field of view is critical)
this is not the case and coding noise is present, with a variance that
is proportional to the portion of the basic pattern not used in coding
a sky position. This also pertains to the WFM. The mask pattern is
larger than the detector, so is never completely recorded. One calls
the portion of the field of view that uses the complete detector the
`fully coded field of view' (FCFOV).  However, in the WFM this is
actually not coded with the complete mask pattern so has coding
noise. Outside the FCFOV only part of the detector is illuminated and
the coding noise becomes worse, although smaller than for a random
mask pattern.  Generally speaking, the local coding noise standard
deviation will be a fraction of the random case that is equal to the
square root of the fraction of the mask pattern employed in coding
that locality. Ergo, we expect the base level of coding noise standard
deviation to be $\sqrt{1-N_{\rm SDD}\,A_{\rm SDD}/A_{\rm mask}}=0.85$
of that for a random mask pattern, with $N_{\rm SSD}=4$ the number of
SDDs, the $A_{\rm SDD}=6.5\,\times\,7.0$~cm$^2$ the sensitive area per
SDD and $A_{\rm mask}=26.0\,\times\,26.0$~cm$^2$ the area of the mask.
The expression under the square root represents the minimum fraction
of the complete mask pattern not used during the coding of a sky
source. Coding noise is a systematic and deterministic noise, meaning
that it does not diminish with exposure time as opposed to statistical
noise. However, one can eliminate or minimize it with IROS.

IROS entails the following critical steps (see also Fig.~\ref{f4}): 
\begin{enumerate}[topsep=0pt,itemsep=0pt,parsep=0pt,partopsep=0pt]
\item
cross correlate the detector image with a mathematical version of the
mask pattern to a sky image, properly normalized that each pixel has
phot s$^{-1}$ cm$^{-2}$ as unit, and an accompanying error image of
the statistical standard deviation in each pixel based on Poisson
photon count statistics \citep[see, e.g.,][]{zand1992};
\item 
identify significantly detected point sources in the cross correlation
image where significance is determined with respect to the local image
standard deviation after a PSF fit that yields their fluxes
\citep[e.g.,][]{zand1994};
\item 
model the detector image for these point sources and exposure time;
\item 
subtract the model detector image from the observed detector image;
\item
cross correlate the residual detector image to a sky image. The sky
image will still have systematic biases because the initial flux
determination is subject to coding noise, but the coding noise will be
smaller than after the previous iteration; 
\item 
go back to step 2 while preserving the original error image and
iterate until there are no more significant point sources remaining;
\item 
add up the PSF models of all iterations and sources back in the last
cross correlation image and perform a PSF fit to determine the
best-estimate fluxes.
\end{enumerate}

The convergence of the iterative process of IROS can be significantly
improved if one takes into account the knowledge about known point
sources in the field of view, in other words uses an X-ray catalog and
the accurate pointing direction of the camera (usually provided by
optical star trackers on board). After all, the known point sources
concern the brightest sources, except for an incidental new transient,
and are the dominant source of coding noise. Furthermore, one can keep
the positions of these frozen during the PSF fits. Freezing the
positions of the bright sources leaves less parameters to fit and
eases the iteration process, particularly in crowded fields (likely in
wide-field applications) because the coding noise can become large. If
there is no pointing information available or its accuracy is
compromised, the camera's X-ray data can be used for plate solving
constellations of X-ray stars and pointing information can be
recovered.

\begin{figure}
\includegraphics[width=1.\columnwidth,angle=0,trim=0cm 0cm 0cm 0cm,clip=true]{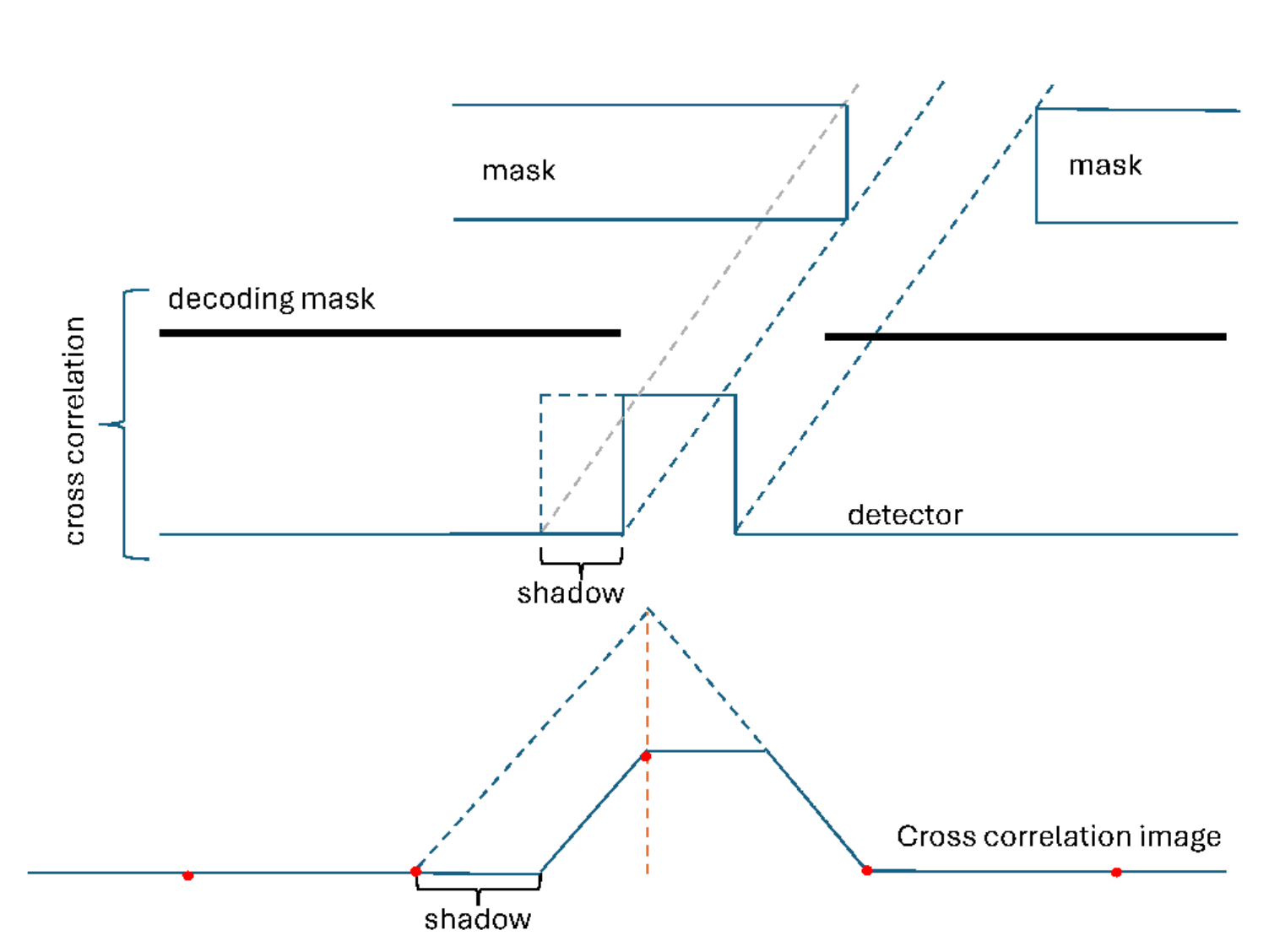}
\caption{PSF formation along the 'fine' direction. In the top section
  of the figure, the rays of an off-axis source illuminate the thick
  mask somewhat sideways and introduce a shadow on one side of the
  projection of the mask element on the detector (middle blue
  graph). When cross correlating with the decoding mask (moving the
  decoding mask from left to right over the detector, see middle blue
  graph), this introduces (bottom blue graph) 1) a flat top to the PSF
  which is off centered; 2) a lower PSF peak representative of less
  sensitive area and 3) a narrower PSF. The red dots indicate where
  the discrete cross correlation would sample the PSF if the sampling
  is equal to the mask element size. The sampled cross correlation
  looks like there is no shadowing but its integral is actually
  smaller.}
\label{f5}
\end{figure}

In IROS, we chose to fit the model to the data in the sky
domain. Alternatively, one could do this in the detector domain, but
in principle this does not matter as translation of detector data to
sky data is the cross correlation procedure which is a linear
combination of detector data into sky data that is independent for
every sky pixel.

\subsubsection{Implementation}

Our implementation of IROS for the WFM has the following specifics.

\begin{itemize}[topsep=0pt,itemsep=0pt,parsep=0pt,partopsep=0pt]
\item
  The WFM data is event-by-event and include positions with a readout
  accuracy that is fraction of the 60 µm fine resolution. We here
  choose to bin these data to detector bins of 0.25 mm$\,\times\,$0.25
  mm, commensurate with the fine mask element size, yielding a total
  detector image size of 632$\,\times\,$632 pixels (filling up dead
  areas between SDDs or along their edges with zeros to obtain paired
  detector images for X and Y cameras) and a sky image size of
  1671$\,\times\,$1671 pixels.
\item
  The cross correlation images of the X and Y cameras are added in
  step 1, resulting in a cross-like PSF which is fitted to
  simultaneously obtain accurate estimates for locations in X as well
  as Y and flux in step 2, while the treatment of the detector images
  of X and Y in steps 3 and 4 remain separate.  Thus, uncertainty
  about the coarse position of each source can be avoided and the
  iterative process is more efficient.
\item
  If, in step 3, the center of the PSF of one source is on the PSF of
  another source, only the most significant source will be included in
  the current iteration. In a next iteration, the ignored source will
  likely be the most significant one.
\item
  If a source has been `subtracted' in a previous iteration, negative
  fluxes are allowed to be subtracted if they are significant. This
  corrects possible over-estimates of the flux in a previous iteration
  (due to a higher coding noise).
\item
  The PSF looks like in Fig.~\ref{f5}. Due the shadowing by the mask
  thickness ($150\,\upmu$m) for an off-axis source, the PSF particularly
  along the fine direction has a plateau that depends on the off-axis
  angle and is off-centered.  Since we have chosen to use 0.25
  mm$\,\times\,$0.25 mm, there is no oversampling which in principle
  introduces ambiguity in source positions, but this will be avoided
  by fixing the point source locations except for unknown sources and
  oversampling in MLM to accurately determine the position of new
  sources.
\end{itemize}

It is noted that IROS can be run independently on the X or the Y
camera.  The sensitivity will decrease by $\sqrt{2}$ and the source
confusion will increase along the coarse direction, but a high level
of redundancy is preserved. Similar for the multiple detectors in one
detector plane.

\subsubsection{Verification}

One step in IROS is the calculation of the model detector image for a
given source configuration (step 3). This is done through a geometric
calculation of the projection of the mask on the discretized detector.
This analytical step enables to verify the self-consistency of IROS:
what one puts in should come out. This implies that source fluxes
should be reproduced exactly and the standard deviation of any
reconstructed sky image after source subtraction should be zero since
no count statistics are involved. We have first tested this with just
a single point source and then in the most complicated sky
configuration possible namely that of the Galactic Center (GC) field.

\begin{figure}
\includegraphics[width=\columnwidth,angle=0,trim=0cm 0cm 0cm 0cm,clip=true]{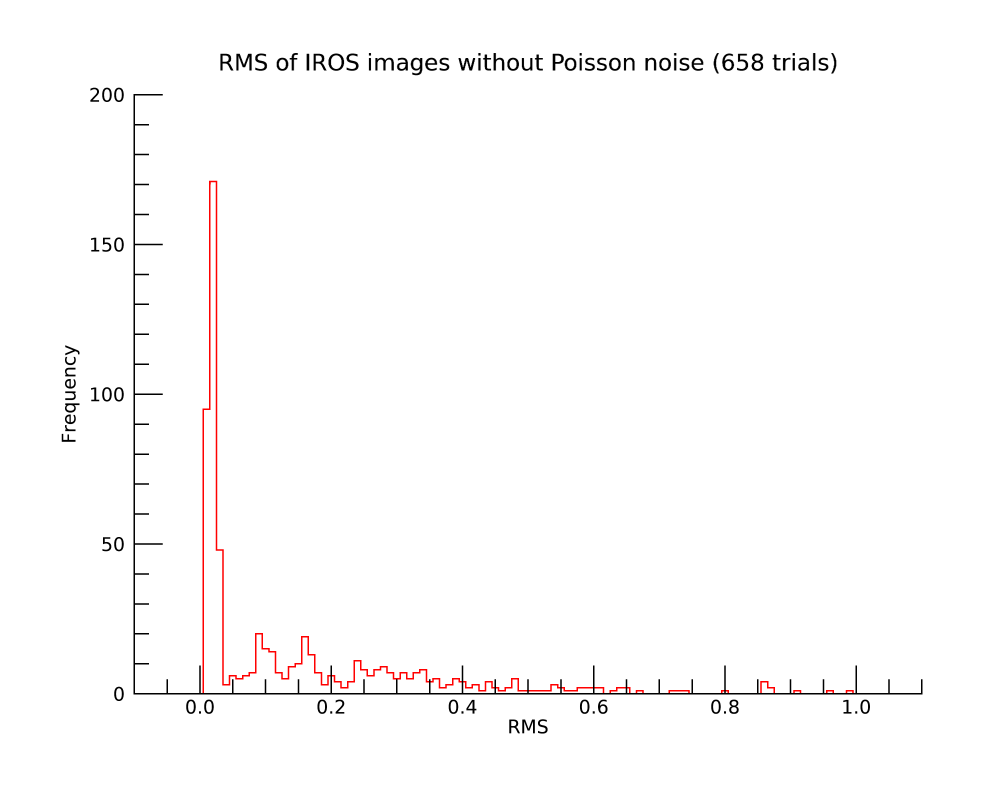}
\caption{Histogram of standard deviation values of significance images
  of the Galactic Center field solution without Poisson noise, for 658
  different pointings (i.e., rotating from 0 to 359 deg in steps of 1
  deg and for stepping diagonally over the image in 299 steps at one
  rotation angle).}
\label{f6}
\end{figure}

The test for the single point source was done in various positions in
the field of view and was every time ideal (standard deviation
zero). For the Galactic Center field this is not the case. We find
that IROS is able to diminish the standard deviation of the full
significance image without point sources to generally 1---10\% of the
value that would be expected from Poisson noise for a 10 ksec
exposure, or equivalent to the Poisson noise for a 1 Msec exposure of
the Galactic Center. Figure~\ref{f6} shows the results in terms of
standard deviation of significance images for 658 different
observation pointing realizations.

\subsection{Maximum Likelihood Method}

\subsubsection{Outline}

Once all sources of detector events have been identified, in other words
once all point and extended sources have been localized and the
backgrounds calibrated, it may become desirable to apply further
diagnostics on the sources: determine the time history of the fluxes
(i.e., determine the light curves) and measure the spectrum, if need be
as a function of time or phase. This implies determination of source
fluxes at many times and photon energies. In principle one can run IROS
for every time and energy bin keeping source positions fixed, but this
will require many computations and long processing times. Basically, this
is because with IROS the flux is calculated in each of the almost 3
million sky pixels (in our WFM application) while one only needs the
flux for around 100 sources. Therefore, one should resort to a method
that is better suited and that method is MLM.

In MLM, the measured detector image $D_i$ (in units of event counts
per detector bin) for a given measured energy band is compared to a
theoretical model consisting of a background component (or multiple
background components, not considered here) and a combination of
$N_{\rm src}$ point sources at different sky locations, all with free
scale factors and taking into account the Poissonian nature of the
counting process. Optimizing these scale factors by maximizing the
likelihood $\mathcal{L}$ gives the source strengths and associated
uncertainties of the involved sources as well as an estimate for the
background plus its uncertainty. In detail, for detector bin $i$ we
can write the expectation value $E_i$ as:

\begin{equation}
E_i = \beta \cdot B_i + \sum_j S_j \cdot M_{ij}
\end{equation}

\noindent
where $M_{ij}$ is the point-spread function at that detector bin $i$
for source $j$, $S_j$ the source flux in sky bin $j$, $B_i$ the
background image in detector bin $i$ and $\beta$ the scale factor of
the total background component. The likelihood $\mathcal{L}$, taking
into account the Poisson statistics, is defined as:

\begin{equation}
\mathcal{L} = \prod_i \frac{E_i^{D_i} \cdot {\rm exp}(-E_i)}{D_i!}
\end{equation}

Assuming for simplicity that we are dealing with one source and one
constant background model $B_i$, ignoring times when the Earth
eclipses part of the field of view, we consider two hypotheses: a) the
zero hypothesis ${\cal H}_0$ stating that the data are described by
background only i.e. \emph{$E^{{\cal H}_0}$} (\emph{$\beta$}) =
\emph{$\beta$}, and we have:

\begin{equation}
\mathcal{L}^{{\cal H}_0}(\beta) = \prod_i \frac{\beta^{D_i}\cdot {\rm exp}(-\beta)}{D_i!}
\end{equation}

\noindent
and b) the alternative hypothesis $\cal H$$_1$ stating that the data
are described in terms of a constant background and a point-source
i.e. \emph{$E^{{\cal H}_1}$} (\emph{$\beta$, $S_1$}) = \emph{$\beta$}
+ \emph{$S_1$} · \emph{M$_{\rm i1}$}, resulting in:

\begin{equation}
\mathcal{L}^{{\cal H}_1}(\beta,S_1) =
\prod_i \frac{(\beta+S_1\cdot M_{i1})^{D_i}\cdot {\rm exp}(-\beta-S_1 \cdot M_{i1})}{D_i!}
\end{equation}

Maximizing the natural logarithm of $\mathcal{L}^{{\cal
  H}_1}$(\emph{$\beta$, $S_1$}) with respect to \emph{$\beta$} and
\emph{$S_1$}, and the natural logarithm of $\mathcal{L}^{{\cal H}_0}$
(\emph{$\beta$}) with respect to \emph{$\beta$}, we can define the
maximum likelihood ratio (MLR) \emph{Q} as

\begin{equation}
Q = -2 \cdot {\rm ln}\left(\mathcal{L}^{{\cal H}_0^{\rm max}}(\beta)/\mathcal{L}^{{\cal H}_1^{\rm max}}(\beta,S_1)\right)
\end{equation}

This (test) statistic is successfully used in high-energy
astrophysics, particularly in the gamma-ray domain, for parameter
optimization and hypothesis testing. More information on the basic
principles, the method and caveats can be found in \cite{Wilks1938},
\cite{Eadie1971} and \cite{Cash1979}, while its application to
high-energy data analysis procedures is described in detail in
\cite{Pollock1981}, \cite{Mattox1996}, \cite{deBoer1992} and
\cite{Abdo2009} for COS-B, CGRO-EGRET, CGRO-COMPTEL and Fermi-LAT,
respectively. An application of the method to BeppoSAX X-ray data is
given in \citet{zand2000}.

In our case the quantity \emph{Q} is distributed as \emph{$\chi^2$}
for one degree of freedom under proper conditions.  Therefore, at
known source locations (e.g., from source catalogues) \emph{Q} values
of 9, 16, 25,... correspond to 3, 4, 5, ...  \emph{$\sigma$} detection
significances.

If we do not know a priori the location of a source we can scan the
sky on a grid by assuming a source location at each grid point and
evaluating the quantity \emph{Q} and so constituting a map - the MLR
significance map. To assess the significance of prominent new-source
features in the MLR-map we have to take into account the number of
trials i.e. the number of independent spatial resolution elements
($\sim$4\arcmin$\times$4\arcmin) searched for in the map, which
degrades the detection significance compared to the assumption of the
one degree of freedom significance at the particular location. The
construction of this map is a tedious and CPU-time consuming process
given the non-linear nature of the optimization procedure and the many
scan locations (preferentially sampled at step sizes smaller than the
angular resolution of the camera) involved to build the map requiring
each time a different detector model. This procedure is feasible as
long it is limited to a small region around a new source for the sake
of its identification, but the computations quickly become
unsustainable for a large field of view (Appendix~\ref{mlmim} shows an
example of a sub-region of the WFM field of view).

\subsubsection{Implementation} 

The practical implementation is as follows. At first, the events in
the detector plane (for a measured energy band) are binned in
30~$\upmu$m by 3.5 mm detector pixels along the fine and coarse
directions, respectively, which is sufficiently small to sample
sufficiently dense the distribution given the 60 \emph{µ}m and 8 mm
FWHM resolution in fine- and coarse directions of the detector. This
results in a 2170$\,\times\,$20 binning mesh per detector plane given
its (X,Y) sizes. This detector image is denoted by \emph{D$_i$} .

The next step is to accurately determine the pixel illumination factor
(PIF) for each pixel in the detector plane for a given source scan
direction. For this purpose we subdivide each detector pixel in
\emph{M} × \emph{M} elements to boost the detector image resolution,
and determine whether the (centers of) these elements are blocked by a
mask element or not when back projecting to that source in the
sky. \emph{M} is typically 3. We perform this procedure for both the
upper and lower mask plane (the mask has a thickness of
150$\,\upmu$m). The current implementation yields a 1 or 0 for an open
and blocked element, respectively. In principle the method can include
energy-dependent absorption effects (the mask is more opaque at lower
X-ray energies) by the Tungsten mask elements using the measured
energy of the events as a proxy for the photon energy, yielding
fractional values between 0 and 1.

We also take into account the averaged penetration depth
\emph{d}$_{\rm pen}$ of the photons in the silicon detector material,
making effectively the detector--mask distance larger by
\emph{d}$_{\rm pen}$. Finally, we blur through convolution of the
detector image, as derived above for the case in which we exactly know
the (X,Y)-location of the absorbed photon in the detector plane, with
a 2D-distribution that accurately describes the blurring of the charge
cloud reconstruction produced by the absorbed photon. This description
is a `modified' sech(x) function (secans hyperbolicus) of the form

\begin{equation}
{\rm modsech}(x\vert \vec{a}) = \frac{\rm a_0} { {\rm cosh} \left( (|\frac{x-{\rm a_1}}{\rm a_2}|)^{\rm a_3} \right) , }
\end{equation}

\noindent
where $x$ is the spatial dependence and $\vec{a}$ a 4-element
parameter array describing the model PSF. Only the `shape' parameters,
a$_2$ and a$_3$, are relevant for the normalized distribution used in
the convolution process.

Thus, we are equipped with the necessary elements, a) the measured
detector \emph{D$_i$}, b) the point spread function \emph{M$_{\rm
  ij}$} for a source at the scan-location and c) a background estimate
assumed to be homogeneous for isotrope diffuse emission, to calculate
the maximum likelihood ratio \emph{Q$_k$} for each scan location
$k$. The square root of this number yields the significance image.

Due to the additive nature of ln$\mathcal{L}$ (i.e., the
multiplicative character of the likelihood $\mathcal{L}$) the method
can easily be extended to analyze an orthogonal camera pair assuming
that both components are identical and operate simultaneously under
similar observation conditions. In that case we can assume that the
same fit (scale) parameters for the models hold for each component of
the camera pair. In practice one should correct for possible slightly
different exposure times and misalignments of each camera.

\section{Simulation approach}
\label{approach}

\subsection{Simulation of sky}

Our baseline for the simulation of point sources is the catalog of
cosmic X-ray point sources as compiled from observations with the
RossiXTE All-Sky Monitor \citep{levine1996,levine2006,levine2011} and
BeppoSAX-WFC \citep{jager1997}. This includes 1.5-12 keV photon fluxes
of those point sources averaged over the 16 year lifetime of the
RossiXTE mission (1996-2012), the best-known positions, often from
optical or radio observations, and average spectra measured with
BeppoSAX-WFC \citep{verrecchia2007}. The fluxes were normalized to the
flux of the Crab and then multiplied with the calculated photon flux
of the Crab expected in the WFM in the bandpass of 2 to 30 keV (2.4
phot s$^{-1}$cm$^{-2}$; based on the Crab spectral model of
\citealt{kirsch2005}). It is noted that this is an educated estimate of
those fluxes with an estimated accuracy of a few tens of percents, but
this is enough for our purposes to assess the WFM imaging capability
and sensitivity.

\begin{figure}
\includegraphics[width=\columnwidth,angle=0,trim=0cm 0cm 0cm 0cm,clip=true]{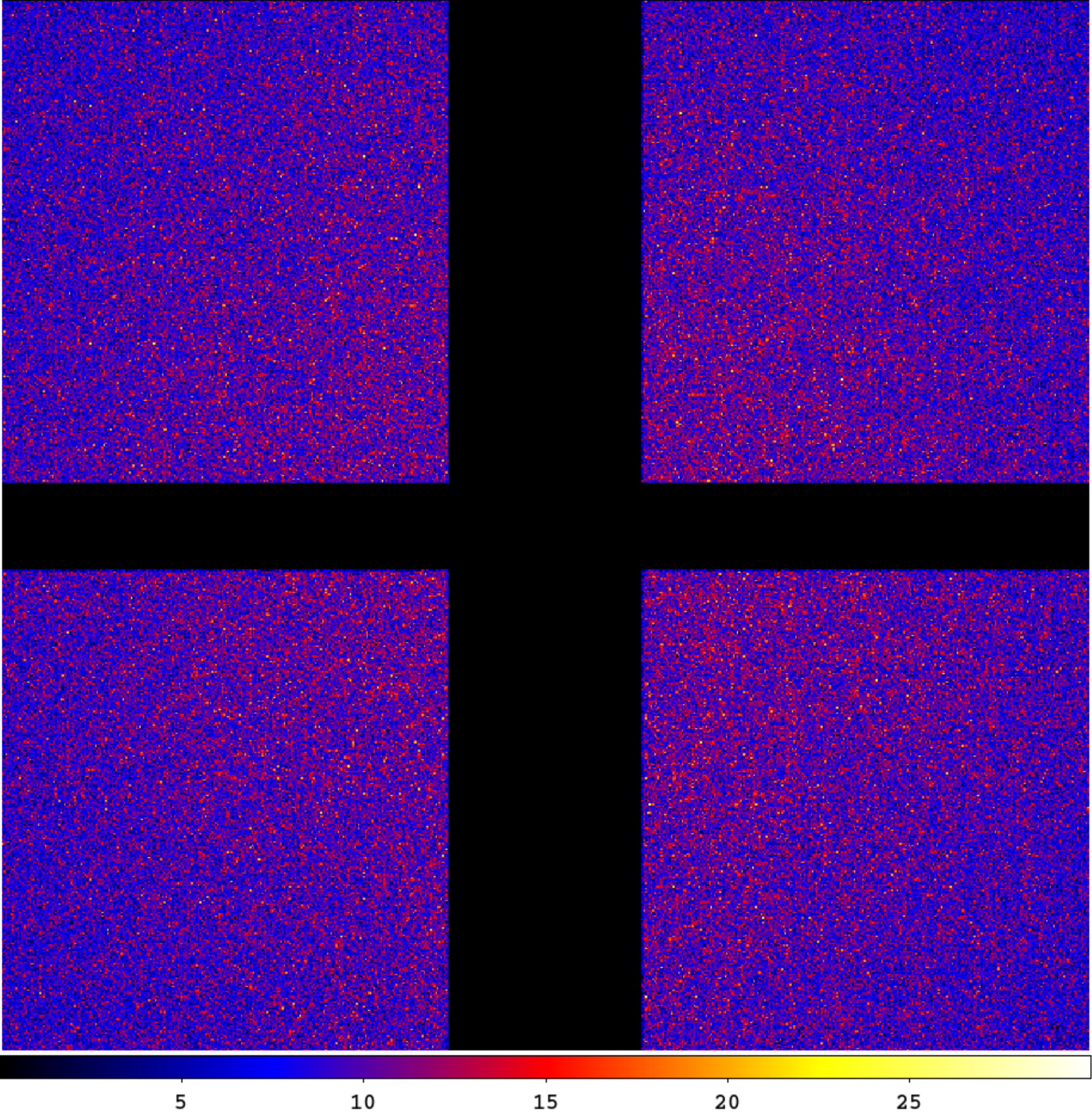}
\caption{Image of X detector (fine resolution along X) for LMC field
  for an observation with an exposure time of 10 ks. Unit is photon
  counts per 0.25 mm$\,\times\,$0.25 mm bin. The color scale bar of
  the pixel values is shown at the bottom. The cross is due to the
  spaces between the four SDD tiles in the detector plane.}
\label{f8}
\end{figure}

We simulate the sky background by a calculation of the detector
response to an isotropic background. This local response of the
detector depends on the field of view as seen from that position on
the detector since the field of view can be modulated by shadowing by
any mechanical structure above it, including the mask. Our simulation
entails a calculation of that field of view and a scaling that is
determined from a Monte Carlo simulation of the cosmic diffuse X-ray
background (CXB) which is 215 phot s$^{-1}$cam$^{-1}$ for the full
bandpass, which is in line with CXB model in
\cite{marshall1980}. Apart from the sky background, real observations
will also include particle-induced background \citep{campana2012} and
the Galactic Ridge emission \citep{valinia1998}. However, the photon
rate of these contributions are expected to be less than 10\% of the
CXB, so negligible at the current level of fidelity.

We employ two sky fields as test cases:

\begin{enumerate}[topsep=0pt,itemsep=0pt,parsep=0pt,partopsep=0pt]
\def\labelenumi{\arabic{enumi}.}
\item
  Faint field, pointed at the Large Magellanic Cloud, in particular
  the point source LMC~X-1. The combined photon count rate of all
  point sources is smaller than the background. The typical flux of a
  bright source in this FOV is a few tens of mCrabs. The brightest
  object is Cen X-3 at 53 mCrab at an off-axis angle of 34\degr. The
  CXB count rate encompasses 97.6\% of the total and so dominates.
\item
  Bright field, pointed at the Galactic Center. The combined photon
  count rate of point sources far exceeds that of the background. In
  particular, this field contains Sco~X-1 at an off-axis angle of
  24\degr, 10\degr\ outside the fully coded FOV. The average count rate
  of Sco~X-1 is 11 times higher than the next brightest source in the
  field, GX~5-1, which is at an off-axis angle of 5\degr. This field is
  an excellent test of systematic errors in the reconstruction. The
  CXB count rate encompasses 19.7\% of the total and, therefore, is
  of secondary magnitude.
\end{enumerate}

\begin{figure}[!t]
\includegraphics[width=\columnwidth]{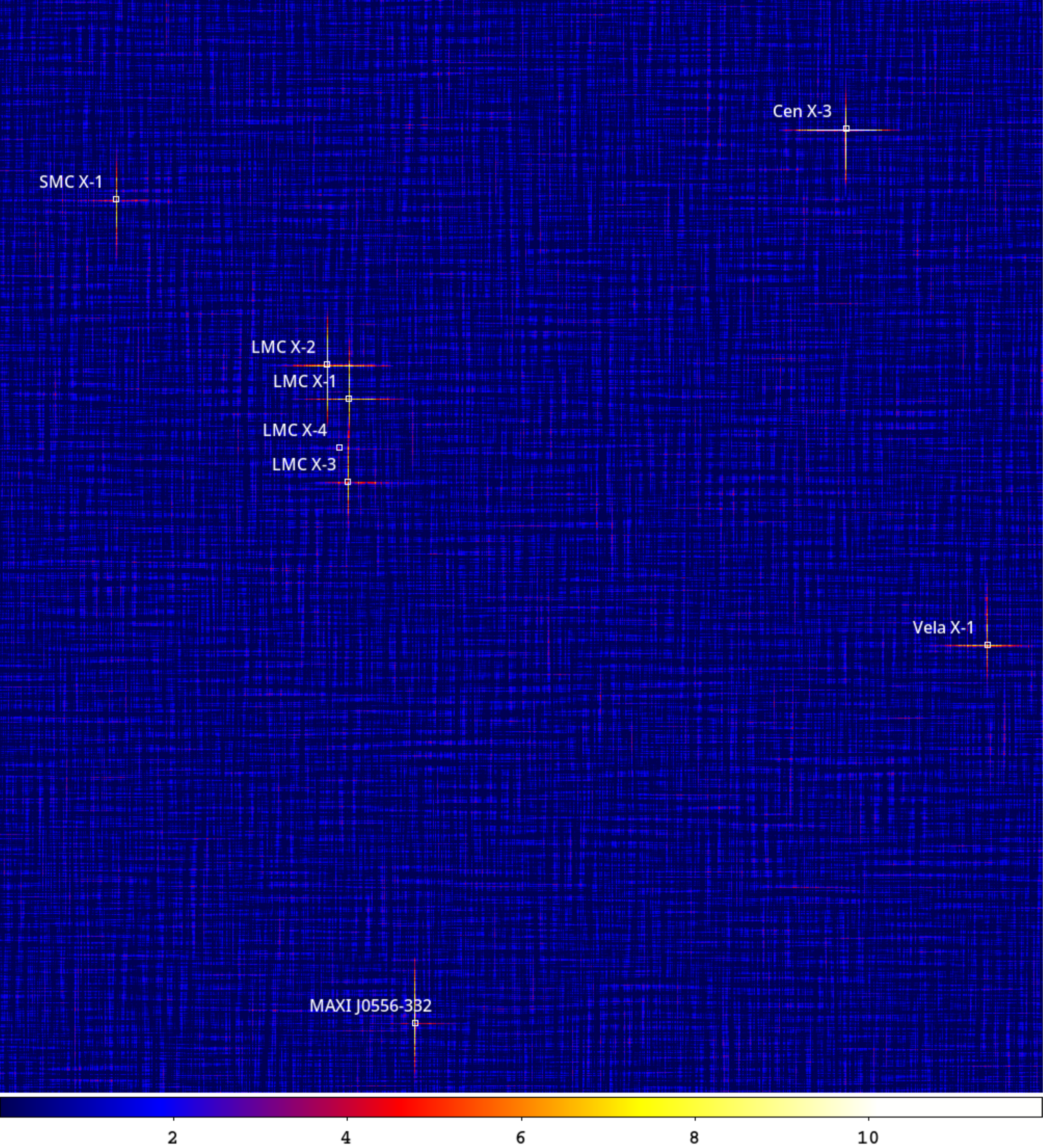}
\caption{Part of the sky image of the LMC observation as obtained
  through IROS, between -28\degr\ and +40\degr\ in the local frame
  along X and -39\degr\ and +25\degr\ in Y, zooming in on 54\% of the
  total image showing the strongest sources in the field of view. The
  pixel unit is significance. Only sources brighter than 13 mCrab are
  labeled. The cross-like PSF of a source is not always apparent
  because its narrowness does not show well at this image
  reproduction; it is clearer when zooming in the pdf version of this
  paper.}
\label{f9}
\end{figure}

\subsection{Simulation of detector response}

Simulations of the WFM detector response are calculated per (array of)
cameras with the physics-based software tool 'WFM Imaging Simulation
Environment for Montecarlo and ANalytical modeling' (WISEMAN). This
tool is comprehensively described in \cite{ceraudo2024}. We here
summarize.

\subsubsection{The simulation process} 

Photons from astrophysical sources (hereafter, primary photons) are
generated in the field of view of the camera starting from the ASM
catalog of X-ray point sources and the diffuse emission of the
CXB. Extended sources such as the Galactic Ridge, supernova remnants
and galaxy clusters are not included. Primary photons (tagged with
celestial coordinates, photon energy and time of arrival) are
projected uniformly on the top surface of the camera, and their
direction is converted to the local frame of reference.

The first loss of primary photons flux occurs above the mask, where
the presence of the Multi-Layer Insulator (MLI) foil is simulated,
with the photons surviving the passage according to a stochastic
process based on the overall transparency of the MLI. Afterward, the
trajectory of each photon is traced through the coded mask and the
path into each (portion of) closed element is recorded, so that the
transmission of each photon may be calculated via a random
process. All surviving photons are then traced to the detector plane,
via a polypropylene foil between mask and detector plane, and only
those whose trajectories intersect with the detectors are followed
further.

\begin{figure}
\includegraphics[width=\columnwidth,angle=0,trim=0cm 0cm 0cm 0cm,clip=true]{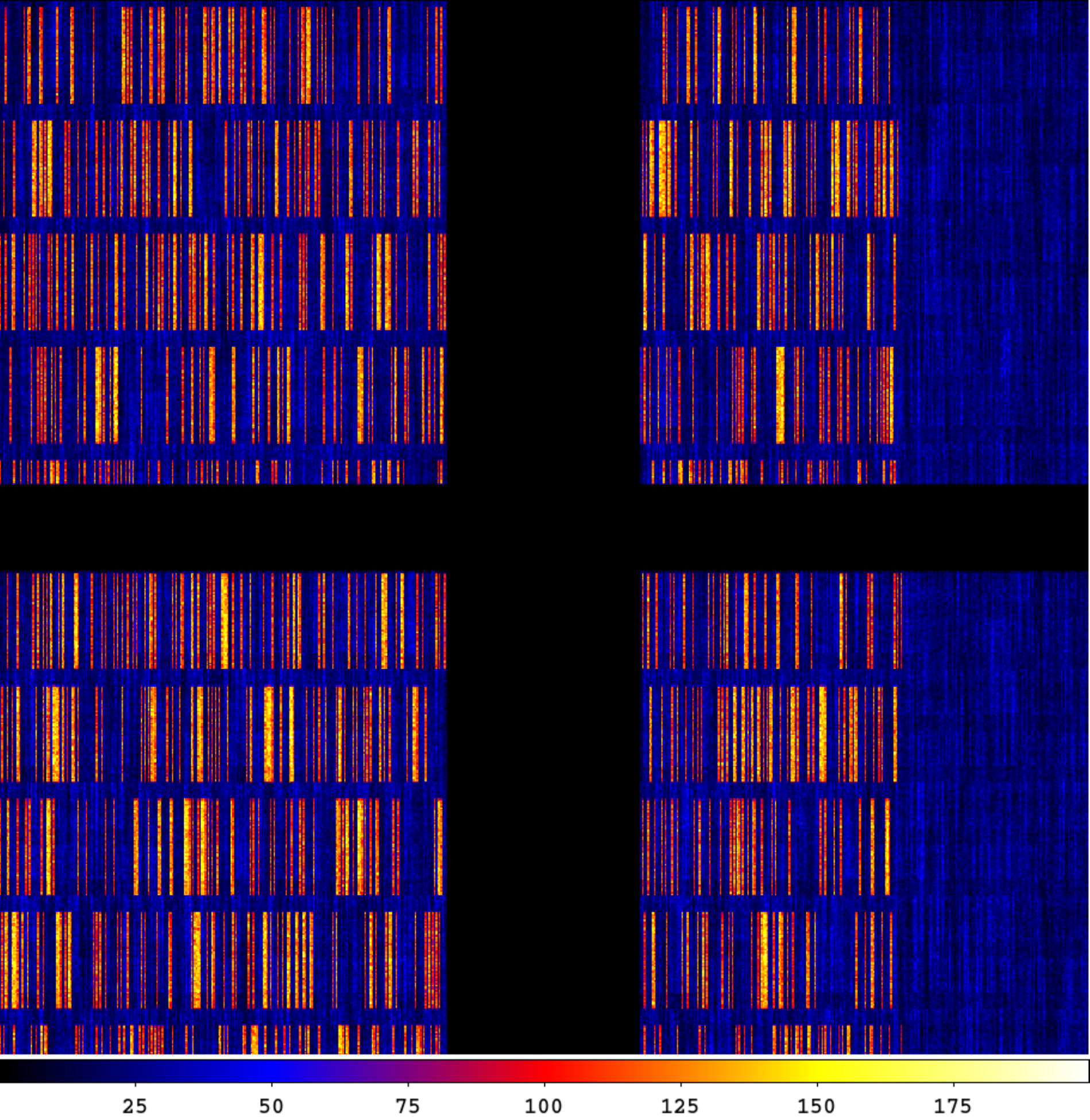}
\caption{Image of X detector for an observation of 10 ks exposure of
  GC field. The illumination by Sco X-1 is obvious. Unit is photon
  counts per 0.25 mm$\,\times\,$0.25 mm bin.}
\label{f10}
\end{figure}

If the photon survives the passage through the inert layers of the top
of the sensor, the interaction of a photon with the sensitive bulk is
evaluated according to its energy and the path it actually traces,
which depends on its direction of approach. If the photon is absorbed
via photo-electric effect in the sensor, its energy is modified by the
Fano noise in silicon, and then the drift of the liberated electrons
is simulated.  During the drift, the initially point-like charge cloud
expands under the effect of diffusion, so that the charge \(\psi_{i}\)
collected by an anode in position \(x_{i}\) is

\begin{equation}
\psi_i = \frac{1}{2} E \left[ 
    {\rm erf}\left(\frac{x_i+p/2-X}{\sigma\sqrt{2}}\right) - 
    {\rm erf}\left(\frac{x_i-p/2-X}{\sigma\sqrt{2}}\right)
                     \right],
\end{equation}

\noindent
where \(p\) is the anode pitch (169 µm), \(X\) and \(Y\) are the
coordinates of the interaction of the photon with the detector, \(E\)
its energy, and \(\sigma\) is the width of the charge distribution due
to diffusion, expressed by

\begin{equation}
\sigma = \sqrt{2Dt+\sigma_0^2} = \sqrt{2\frac{k_{\rm B}T}{q}\upmu\frac{Y}{\upmu\epsilon}+\sigma_0^2} = \sqrt{2Y\frac{k_{\rm B}T}{q\epsilon}+\sigma_0^2},
\end{equation}

\noindent
with $D$ the diffusion coefficient, $t$ the drift time, \(\sigma_{0}\)
the initial width of the charge cloud, \(T\) the detector temperature,
$\upmu$ the electron mobility, $\epsilon$ the electric field, \(q\) the
elementary charge and \(k_{B}\) Bolzmann's constant
\citep{campana2011}.

Readout is implemented by adding noise to each anode reading,
converting the continuous energy values to the integer units of the
analog-to-digital converter and adding channel-wise offsets as well as
an event-dependent baseline.

The final step of the simulation essentially consists of estimating
the parameters of the interacting photon (\(X\), \(Y\) and \(E\)) from
the recorded list of digitized anode readings. This procedure involves
the application of user-provided calibration tables (e.g., gain,
offset etc.) as well as a fit of the recorded charge profile.

\begin{figure}[!t]
\includegraphics[width=\columnwidth]{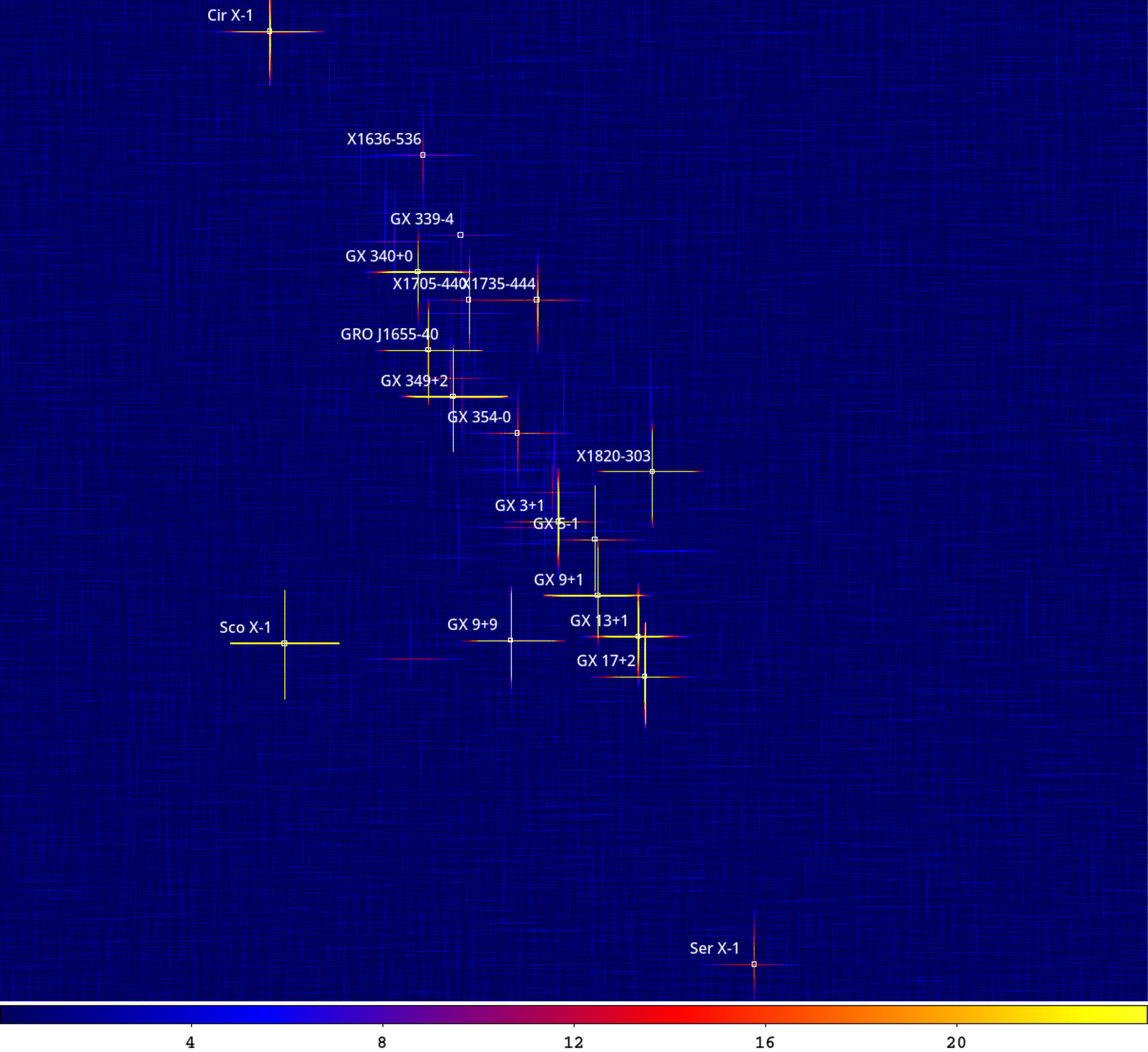}
\caption{Part of the sky image of the GC observation as obtained
  through IROS, between -44\degr\ and +45\degr\ in in the local frame
  along X and -35\degr\ and +35\degr\ in Y, zooming in on 30\% of the
  total image showing the strongest sources in the field of view. The
  pixel unit is significance. Only sources brighter than 80 mCrab are
  labeled. The cross-like PSF of a source is not always apparent
  because its narrowness does not show well at this image
  reproduction; it is clearer when zooming in the pdf version of this
  paper.}
\label{f11}
\end{figure}

At the end of the simulation, an output file resembling the actual data
stream of an operating instrument is provided, featuring the list of the
inferred energies and impact (X,Y) positions of the photons.

\begin{figure*}
\includegraphics[width=2\columnwidth]{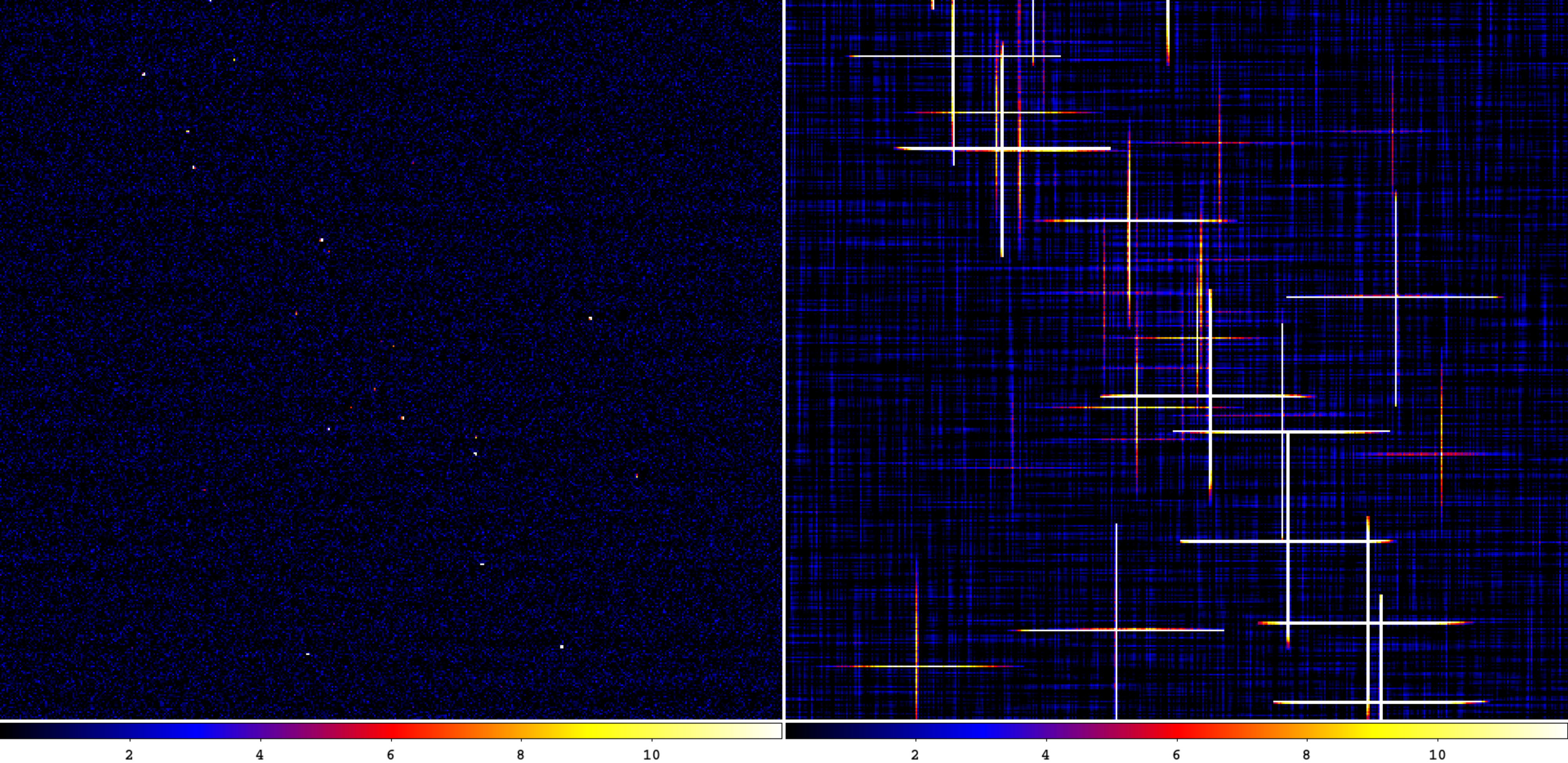}
\caption{Comparison of sky reconstructions of the Galactic Center
  field between the 2$\,\times\,$1.5D configuration (right) and the
  1$\,\times\,$2D configuration (left), zooming in on central quarter of
  the field of view.}
\label{f13}
\end{figure*}

\subsubsection{Strengths and limitations} 

The WISEMAN simulator can be employed to estimate the scientific
performances of an arbitrary arrangement of WFM cameras, hence it is
not limited to a specific configuration (e.g., LOFT, eXTP, STROBE-X,
ARCO or LEM-X). Moreover, since nearly every camera subsystem is
implemented, parametric simulations can be performed to perfect the
design as a function of geometry, building materials and general
characteristics and performances of the various components
(e.g. electronic noise, defective readout channels).

Deviations from the ideal geometric configuration may also be
investigated, since each subsystem may be freely positioned around its
baseline location, allowing to simulate three-dimensional shifts and
rotations of mask and detectors, and therefore their impact on the
scientific performances of the instrument. Likewise, the sensitivity of
the reconstruction algorithms to deviations from the baseline
configurations that are unaccounted for may be assessed in this way.

At the current stage of development, the simulator features some
limitations. For example, all primary photons (and associated detected
events) are treated completely independently from each another,
implying the impossibility to quantify such effects as pile up and
dead time although that is expected to be limited. Additionally, only
photons from astrophysical sources are simulated and their energy and
direction stay constant until the final interaction with the detector:
no particle background or particle-induced photon background are
implemented, and primaries are not reprocessed via scattering.
Likewise, events deriving from electronic noise are not considered.
Finally, as previously mentioned, the astrophysical photons sources
are currently limited to point sources, provided via a catalog, and
the CXB.  No Galactic Ridge has been included so far.

Also, at the current stage of development, the model that is used for
simulating input for the IROS tests assumes infinitely dense masks and
detectors and real photon absorption positions. The algorithm to
derive photon detector positions is not developed yet at a
sufficiently high fidelity to avoid large-scale low-amplitude features
in detector images that are noticeable in IROS images. For the MLM
simulations, the model is sufficient and avoids these idealizations,
because the focus there is on (spectra of) bright sources.

\section{Image simulations}
\label{sims}

\subsection{IROS on LMC field}

The detector image of the WFM X camera of the simulated 10 ksec
observation on the LMC is shown in Fig.~\ref{f8}. This simulation
involves 2.7 million photons. It shows the following features:

\begin{itemize}[topsep=0pt,itemsep=0pt,parsep=0pt,partopsep=0pt]
\item
  a thick vertical center bar running and a thin horizontal one. This
  aligns with the spaces between the 4 SDDs. The space is larger in X
  because there is more hardware along those edges of the SDDs;
\item
  no obvious exposures of point sources, because their count rates are
  negligible with respect to that of the CXB;
\item
  a large-scale non-uniformity peaking at the center of the detector.
  This is due to the field of view as seen from the center of the
  detector being approximately 20\% larger than from the corners.
\end{itemize}

The Y camera shows a similar detector image (not shown) but rotated
by 90\degr.

The IROS solution of this observation is shown in Fig.~\ref{f9},
zooming in to 41\% of the total image with the 6 most significant
sources. The weakest source is LMC X-4 (not shown) at a significance
of 3.1, suggesting a 5-sigma detection limit of 7 mCrab for this
field.

\subsection{IROS on GC field}

Figure~\ref{f10} shows the detector image of the X camera for the 10
ksec simulated observation of the Galactic Center. This measurement
with the X camera includes 11 million photons, 21\% from the CXB and
44\% from Sco X-1 -- the brightest point source in the field of
view. The Y camera image (not shown) is similar. Many of the 163
catalog sources in the field of view leave photons on the
detector.

This observation is in juxtaposition to the one on the LMC. It is
dominated by the brightest point sources and as such represents the
least sensitive field in the sky. This is clear from the detector
image which is dominated by the illumination by Sco X-1 at an off axis
angle of 24\degr. This illumination also clearly shows a shadowgram of
the coded mask with the 14 mm long mask elements and the 2.4 mm ribs
between them along the same axis. The CXB is not discernable, as
opposed to the image of the LMC field (Fig.~\ref{f8}).

In Fig.~\ref{f11}, the IROS solution is shown for this
observation. This 10 ksec observation reveals detections of 41 point
sources. IROS is effective in that it brings down the overall standard
deviation in the significance image down from 28 to 1.1, the large
initial standard deviation being due to coding noise from mostly Sco
X-1. After the first IROS iteration, involving only the subtraction of
Sco X-1, the standard deviation already goes down to 6. The
significance of Sco X-1 is very high at about 1400. The significance
of a source very close to the Galactic center, 1E1747.0-2942, is
2.3. This translates to a 5-sigma sensitivity of 25 mCrab, roughly 3
times less sensitive than the faint LMC field which, therefore, shows
the dynamic range of sensitivities for any WFM observation within the
same exposure time.

Sco X-1 is 24\degr\ from the Galactic Center. With a single camera
pair field of view of 90$\,\times\,$90 square degrees it is cumbersome
to keep Sco X-1 outside the field of view while keeping focused on the
Galactic Center. It would imply that the Galactic Center needed to be
placed some 25 degrees from the edge of the field of view, or 20-30
degrees from the center, depending whether it would be moved towards a
corner or not. Since the FCFOV is the central 29$\,\times\,$29 square
degrees part of the camera field, the loss in detector area covering
the Galactic Center would be limited to 15-30\%.

The detrimental effect of Sco X-1 on the sensitivity was calculated by
doing the same GC simulation while `turning off' Sco X-1. This shows a
40\% improvement in the sensitivity of the 10 ksec GC observation for
the on-axis position and less for far away from Sco X-1. Because Sco
X-1 is outside the FCFOV, there are parts in the sky image opposite to
where Sco X-1 resides that are not affected by the presence of Sco
X-1, but the vast majority of the Galactic Bulge sources are within
reach of the Sco X-1 cross talk.

In Appendix~\ref{3wfm} we show, as illustration, a simulation of a
complete WFM instrument for the eXTP configuration (N=3).

\subsection{Special test: imaging with 1$\,\times\,$2D instead of 2$\,\times\,$1.5D}

In order to assess how the WFM 2$\,\times\,$1.5D configuration
compares to a 1$\,\times\,$2D configuration with similar capabilities,
we set up a simulation with a 2D camera that has twice the detector
area, but identical field of view and angular resolution as the
2$\,\times\,$1.5D configuration.  This implies a simple scaling of all
linear sizes by a factor of $\sqrt{2}$. The result is shown in
Fig.~\ref{f13}. Essentially, the significances of all sources are
similar as are the positional uncertainties for non-cataloged sources,
but the source confusion in the 2$\,\times\,$1.5D configuration is
substantially worse if two sources are located on each other's PSF
cross which extends over a $\sim$256 times larger part of the field of
view. This is particularly relevant for faint transients in the the
Galactic Bulge.

\section{Spectral simulations}
\label{specsims}

While IROS is most effective in image processing and will provide a
complete list of point sources detected in the field of view with flux
inferences at an accuracy of up to typically a few percent
\citep[e.g.,][]{goldwurm2022}, MLM, although also capable of
producing images but at high CPU cost (see Appendix~\ref{mlmim}),
excels in the detailed analysis of those sources, in particular the
determination of the high-resolution time history of their flux and
their spectrum. We here show the spectral capability of MLM. One of
the sources in Galactic Center field, GRO J1655-40, is simulated with
the following spectral ingredients that are inspired by the spectra of
black hole X-ray binaries in their hard state \cite[e.g., review
  by][]{kalemci2022}:

\begin{itemize}[topsep=0pt,itemsep=0pt,parsep=0pt,partopsep=0pt]
\item
  A continuum consisting of a power law with a photon index of 1.86
  and a normalization of 1.14 phot s$^{-1}$cm$^{-2}$keV$^{-1}$ at 1
  keV. This implies a brightness of 200 mCrab for this source,
  ignoring the line contributions below which are substantially
  smaller.
\item
  Interstellar absorption with \emph{N}$_{\rm
    H}=0.61\,\cdot\,10^{22}$ cm$^{-2}$, through the photoionization
  cross sections of \cite{verner1995} and \cite{verner1996} and the
  abundances of \cite{anders1989}
\item
  A narrow Gaussian emission line at 6.4 keV with an intensity of
  0.00263 phot s$^{-1}$cm$^{-2}$. This line has an equivalent width of
  60 eV.
\item
  A Laor emission line profile at 6.91 keV from an accretion disk
  surface with an emissivity dependence that is proportional to
  $R^{-3}$, an inner radius of 1.235 G\emph{M}/c$^2$, an outer radius
  of 400 G\emph{M}/c$^2$, an inclination of the disk to the line of
  sight of 30 degrees and an intensity of 0.0189
  phot\,s$^{-1}$cm$^{-2}$. This line has an equivalent width of 600
  eV.
\end{itemize}

\begin{figure}
\includegraphics[height=\columnwidth,angle=270,trim=0 1.3cm 0 0,clip=true]{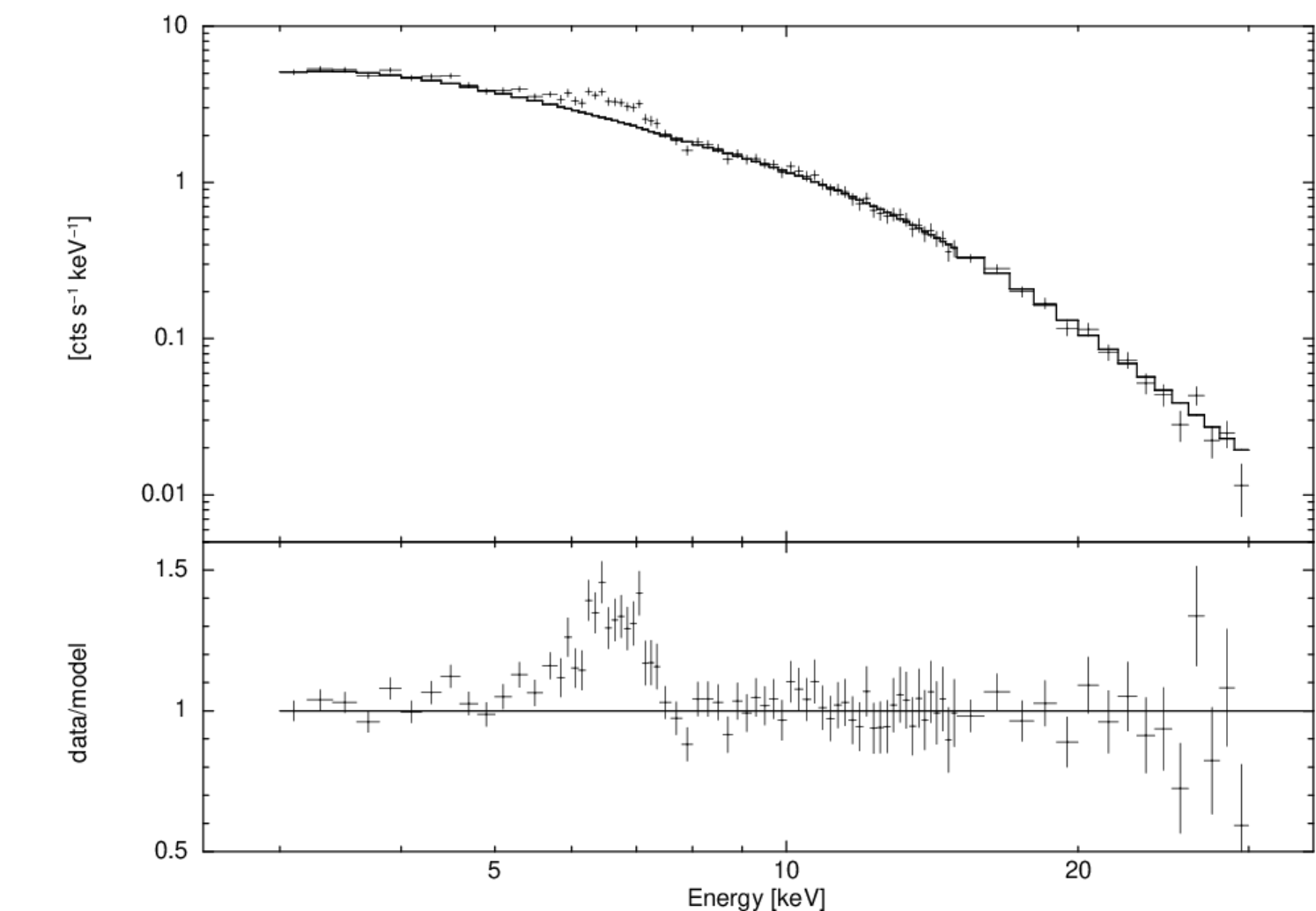}
\caption{Top panel: MLM reconstruction of the spectrum of GRO J1655-40
  (data points) and best-fit model spectrum (drawn histogram). For
  details, see main text. Bottom panel: comparison between data points
  and model, after setting the normalizations of the two spectral
  lines to zero.}
\label{f14}
\end{figure}

\noindent
We applied MLM to reconstruct the spectrum for GRO J1655-40 and model
it with the 4 components above, using XSPEC version 12.15.1
\citep{arnaud1996} with the public response functions for the WFM. The
result is shown in Fig.~\ref{f14}. We were able to reproduce the free
model parameters as follows: $N_{\rm H}=(1.63\pm0.44)\cdot
10^{22}$~cm$^{-2}=+2.3\sigma$, $\Gamma=1.90\pm0.03=+1.2\sigma$,
PL-normalization=$1.268\pm0.104$~phot~s$^{-1}$cm$^{-2}$keV$^{-1}=+1.2\sigma$,
narrow Gauss line $E=6.40\pm0.08$~keV=$0.0\sigma$, narrow Gauss line
intensity is $0.0026\pm0.0008=-0.1\sigma$, Laor line
$E=6.90\pm0.06$~keV=$-0.1\sigma$, Laor line intensity is
$0.0238\pm0.003$~phot~s$^{-1}$cm$^{-2}=+1.6\sigma$. Therefore, MLM
reproduces the multi-component spectrum of this 200 mCrab source
accurately.

\section{Conclusion and future work}
\label{conc}

We have developed and verified the performance of three independent
software packages for application in the WFM: WISEMAN, to simulate the
instrument complete signal chain from celestial X-ray sources to event
data streams; IROS, to reconstruct iteratively and fast the sky image
from the coded detector data for a particular observation; MLM, to
quickly determine the fluxes for all X-ray sources in the field of
view for any time interval within the observation and any photon
energy interval.  We find consistent results throughout which support
the validation of all three software packages.

Experimentation with the software has shown that the concept of a pair
of identical orthogonal 1.5D coded aperture cameras performs equally
well as one 2D coded cameras of the same fine angular resolution, with
the same field of view and an equal detector area as both 1.5D cameras
combined, except for faint transients in the Galactic Bulge. The
advantages of this 2$\,\times\,$1.5D setup are that it is possible to
use light-weight low-power consumption SDDs and that some level of
redundancy is built in. The disadvantage is that the larger cross-like
PSF introduces an increased level of source confusion and cross talk
between sources when they are on each other's PSF, but at the
brightness range that such a camera would be active the source
confusion is not an important issue except perhaps in the Galactic
Center region.

Our work confirms end-to-end the feasibility of the WFM concept. Its
modularity makes it easily adaptable to any mission/observatory concept
and low-risk due to the implicit redundancy.

The simulation software enables to test different configurations of
the WFM and mask pattern, and the vulnerability to imperfections such
as slight misalignments between SDDs in one camera, between mask and
detector plane and between cameras in a pair. Regarding the latter,
the independent functionality of a single camera comes in
handy. Future work will focus on the verification of the alignment
budget, experimentation with the mask pattern resolution to align with
improvements in SDD spatial resolution, with timing resolution and
with spectral resolution in the signal chain.

\begin{acknowledgements}
This work was performed by the eXTP-WFM Simulations Working Group that
was active in 2020-2024 until the WFM instrument was removed from
eXTP. Support was provided by INAF Rome, SRON and ICE-CSIC
Barcelona. The employed software builds on earlier software written
for missions BeppoSAX, AGILE, CGRO and INTEGRAL. We acknowledge all
colleagues that were involved in writing the code. It is noted that no
AI tool has been employed in the coding, data analysis or manuscript
preparation.
\end{acknowledgements}

\bibliographystyle{aa} \bibliography{aa59395-26_references}

\begin{appendix}

\nolinenumbers
\onecolumn

\section{MLM on GC field}
\label{mlmim}

\begin{figure}[ht]
\center{
\includegraphics[width=0.9\columnwidth]{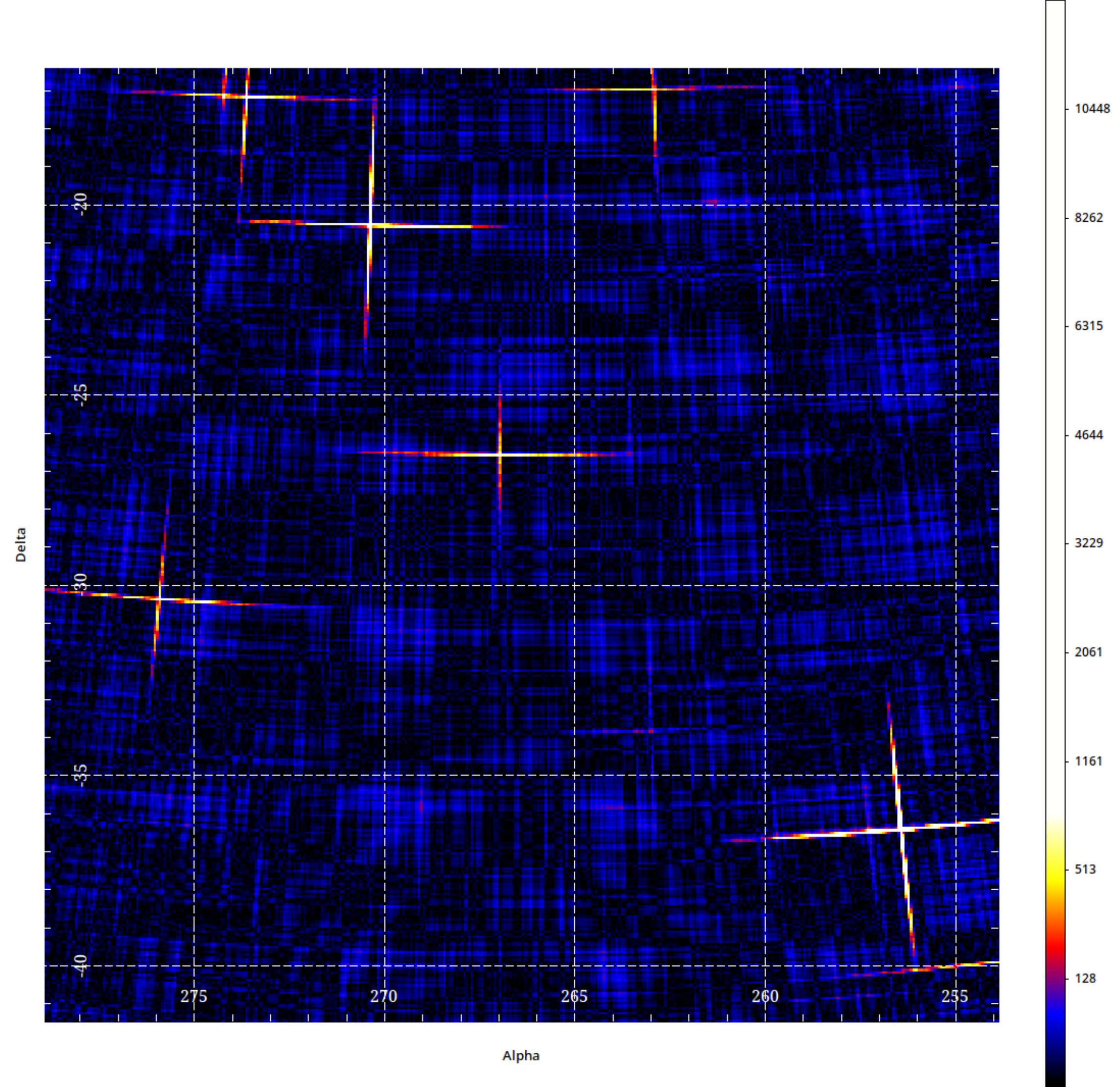}
}
\caption{MLM solution of the central 25$\,\times\,$25 square degrees
  of the GC observation, projected on an R.A.-Decl. plane
  (Eq. 2000.0). Sources are 4U 1820-30 at (R.A., Decl.)=(275.9,-30.4),
  GX 9+1 at (270.4,-20.5), GX 3+1 at (267.0,-26.5), GX 9+9 at
  (262.2,-17.0), GX 13+1 at (273.6,-17.2), GX 354-0 at (263.0,-33.8),
  GX 349+2 at (256.4,-36.4) and partially GRO J1655-40 at
  (353.5,-39.8). GX 5-1 is missing at (270.3,-25.1) because it
  is included in ${\cal H}_0$.  The pixel scale is Q (proxy for
  significance).}
\label{f12}
\end{figure}

MLM can be employed to generate images. An example is shows in
Fig.~\ref{f12}, for part of the GC sky image. For this exercise, GX
5-1 was included in H0 and the image was generated by scanning the
field of view for deviations from H0. As said, this is a
computationally intensive exercise. The generation of this image,
which covers only 8\% of the field of view, took approximately 24 hr
on a Linux Debian system with a 12th generation Intel(R) Core(TM)
i7-12700K CPU.

\section{Complete eXTP-WFM simulation}
\label{3wfm}

\begin{figure}
  \centering
  \includegraphics[width=\linewidth]{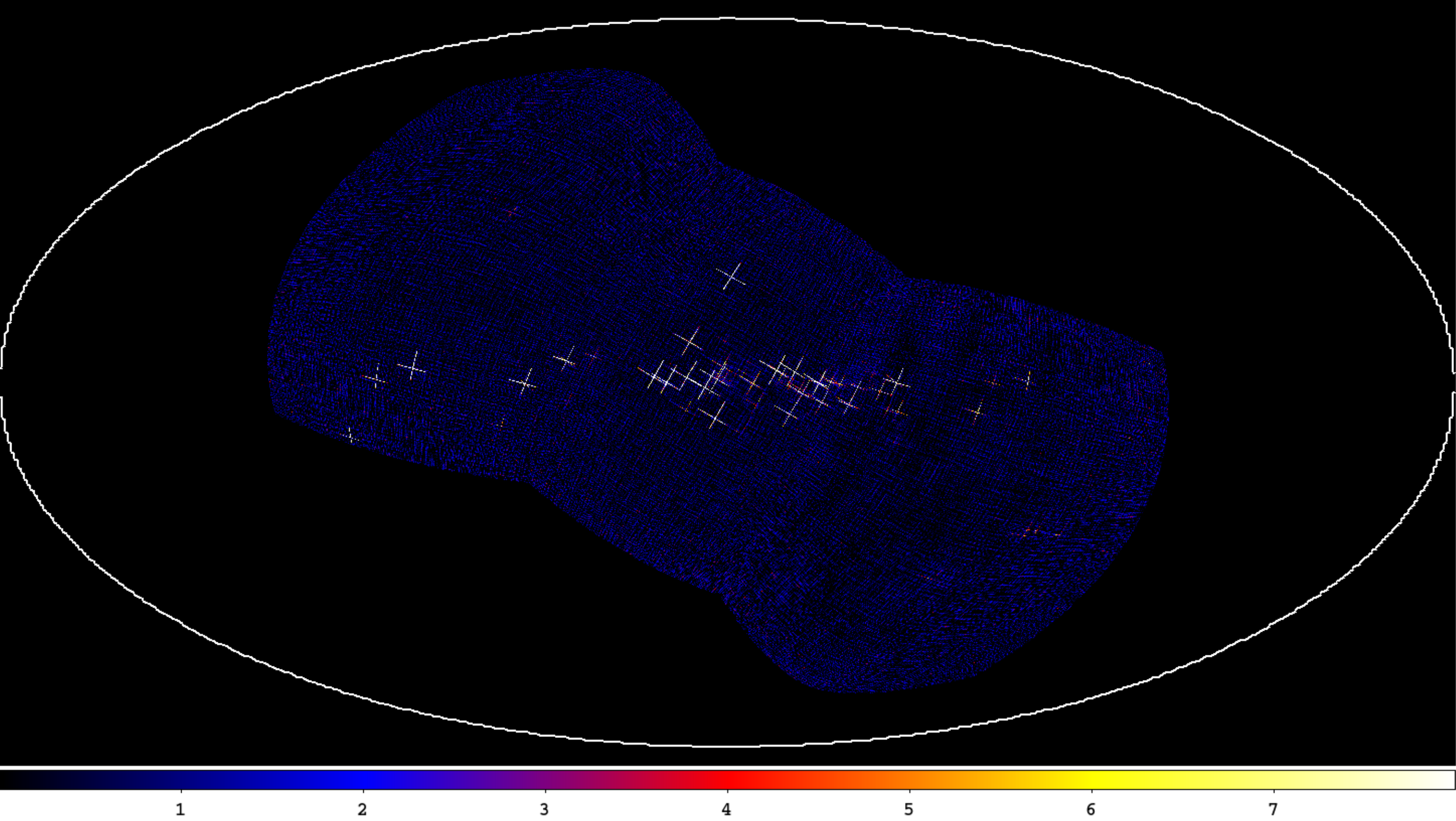}
  \makebox[0pt][r]{
    \raisebox{7cm}{%
      \includegraphics[width=.3\linewidth]{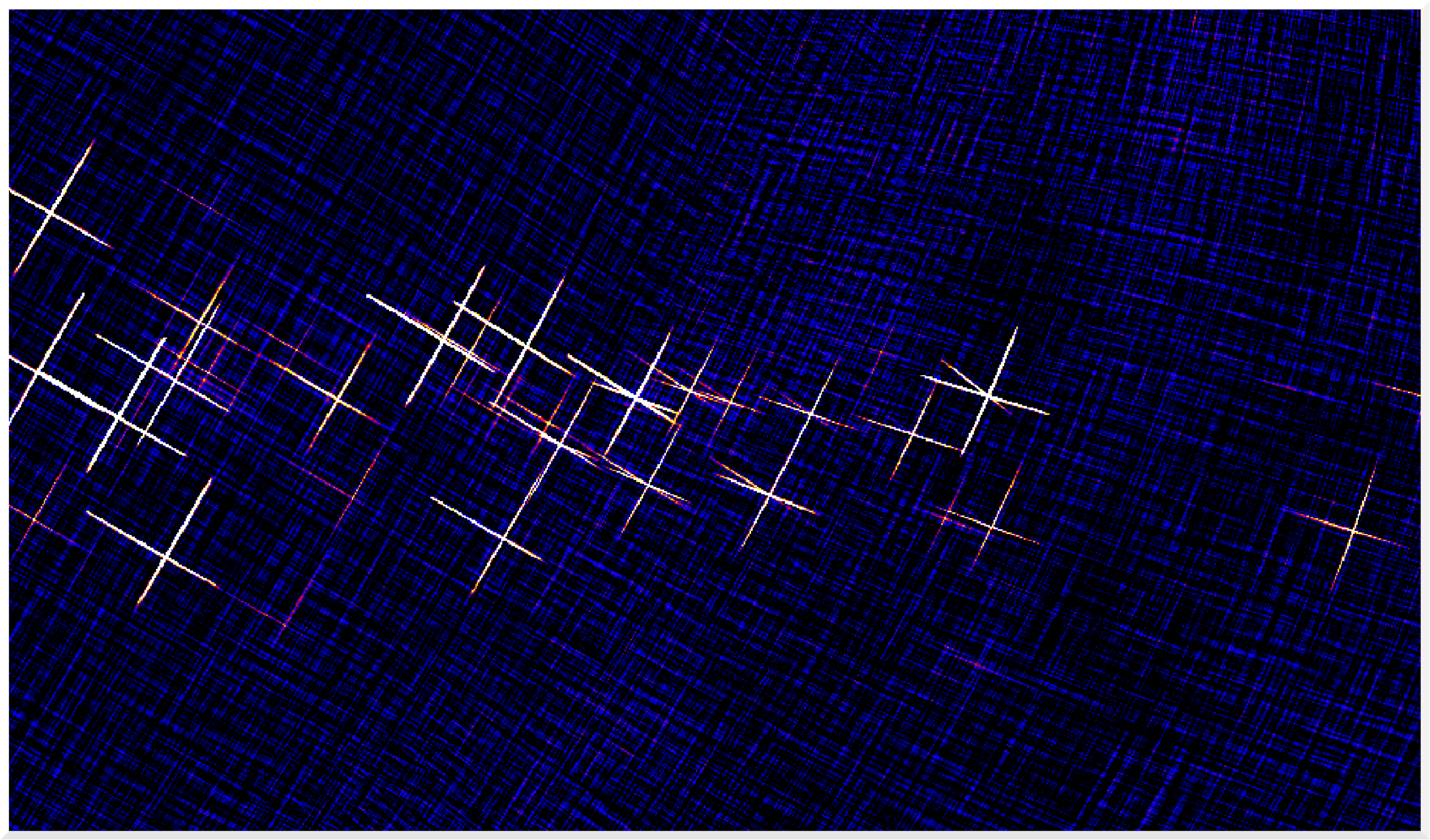}
    }\hspace*{1em}%
  }%
\caption{Example of an IROS solution of a 10 ksec eXTP-WFM (N=3)
  observation of GX 3+1 for the Sun position on March 15, mapped in an
  Aitoff projection in Galactic coordinates. The color scale is mapped
  between 0 and 8 significance. The inset shows a zoomed-in region
  where the field of views of two camera pairs overlap.}
\label{f7}
\end{figure}

In order to show what the potential is of a complete WFM instrument,
we have simulated an observation centered close to the Galactic Center
(i.e., GX 3+1) for the Sun position on March 15 for 3 camera pairs as
would have been applicable for eXTP. The IROS solution is shown in
Fig.~\ref{f7}. This single observation covers the Galactic plane
between -100 and +100 degrees in Galactic longitude and a substantial
range of latitudes above and below the plane. It covers 40\% of the
sky (5.0 sr). It also covers 75\% of all ASM sources and bright
Galactic X-ray sources and 83\% of all bright ASM sources (i.e., with
an average flux higher than 10 mCrab). There are a few sources which
are in overlapping FOVs of 2 adjacent camera pairs. These show double
crosses at a small angle. Thus, the source confusion increases
somewhat in these overlap regions but is partly compensated by an
increased sensitivity.

\end{appendix}

\end{document}